\documentclass[cmc,article,accept,moreauthors,pdftex]{Definitions/tsp} 
\usepackage[normalem]{ulem} 
\usepackage{amsfonts} 
\usepackage{mathcomp} 
\usepackage{CJKutf8} 
\usepackage{pifont} 
\usepackage{bm} 
\usepackage{bbm} 
\graphicspath{{./Definitions/}} 

\makeatletter
\def\T@n@@nc@d@ngM@cr@M@d{}
\def\LY@n@@nc@d@ngM@cr@M@d{}
\makeatother

\let\orignewcommand\newcommand  
\let\newcommand\providecommand  
\usepackage{verse}
\let\newcommand\orignewcommand  

\newsavebox\foobox

\renewcommand{\dagger}{\mathchar"2279}
\renewcommand{\ddagger}{\mathchar"227A}

\newcommand{\mmathit}[1]{
  \ifthenelse{\equal{#1}{\ln}}{\mathit{ln}}{
    \ifthenelse{\equal{#1}{\max}}{\mathit{max}}{\mathit{#1}}
  }
}
\makeatother
\robustify{\footnote}

\DeclareUnicodeCharacter{1E45}{\.{n}}
\DeclareUnicodeCharacter{1E41}{\.{m}}
\DeclareUnicodeCharacter{2003}{\quad}
\DeclareUnicodeCharacter{2009}{\thinspace}
\DeclareUnicodeCharacter{2002}{\enspace{}}
\DeclareUnicodeCharacter{2005}{\thinspace}
\DeclareUnicodeCharacter{0263}{\textipa{G}}
\DeclareUnicodeCharacter{A0}{~}
\DeclareUnicodeCharacter{2460}{\textcircled{\scriptsize{1}}}
\DeclareUnicodeCharacter{2461}{\textcircled{\scriptsize{2}}}
\DeclareUnicodeCharacter{2462}{\textcircled{\scriptsize{3}}}
\DeclareUnicodeCharacter{2463}{\textcircled{\scriptsize{4}}}
\DeclareUnicodeCharacter{2464}{\textcircled{\scriptsize{5}}}
\DeclareUnicodeCharacter{2465}{\textcircled{\scriptsize{6}}}
\DeclareUnicodeCharacter{2466}{\textcircled{\scriptsize{7}}}
\DeclareUnicodeCharacter{2467}{\textcircled{\scriptsize{8}}}
\DeclareUnicodeCharacter{2468}{\textcircled{\scriptsize{9}}}
\DeclareUnicodeCharacter{2070}{\textsuperscript{0}}
\DeclareUnicodeCharacter{2074}{\textsuperscript{4}}
\DeclareUnicodeCharacter{2075}{\textsuperscript{5}}
\DeclareUnicodeCharacter{2076}{\textsuperscript{6}}
\DeclareUnicodeCharacter{2077}{\textsuperscript{7}}
\DeclareUnicodeCharacter{2078}{\textsuperscript{8}}
\DeclareUnicodeCharacter{2079}{\textsuperscript{9}}
\DeclareUnicodeCharacter{02C2}{<}
\DeclareUnicodeCharacter{2033}{\relax\ifmmode '' \else $''$\fi}
\DeclareUnicodeCharacter{2034}{\relax\ifmmode ''' \else $'''$\fi}
\DeclareUnicodeCharacter{2026}{\relax\ifmmode … \else $\ldots$\fi}
\DeclareUnicodeCharacter{0229}{\c{e}}
\DeclareUnicodeCharacter{016F}{\r{u}}
\DeclareUnicodeCharacter{127}{\relax\ifmmode\rm\hbar\else $\rm\hbar$\fi}
\DeclareUnicodeCharacter{3AC}{\relax\ifmmode\acute{\alpha}\else $\acute{\alpha}$\fi}
\DeclareUnicodeCharacter{3AD}{\relax\ifmmode\acute{\varepsilon}\else $\acute{\varepsilon}$\fi}
\DeclareUnicodeCharacter{3AE}{\relax\ifmmode\acute{\eta}\else $\acute{\eta}$\fi}
\DeclareUnicodeCharacter{3AF}{\relax\ifmmode\acute{\iota}\else $\acute{\iota}$\fi}
\DeclareUnicodeCharacter{3CC}{\relax\ifmmode\acute{o}\else $\acute{o}$\fi}
\DeclareUnicodeCharacter{3CD}{\relax\ifmmode\acute{\upsilon}\else $\acute{\upsilon}$\fi}
\DeclareUnicodeCharacter{3CE}{\relax\ifmmode\acute{\omega}\else $\acute{\omega}$\fi}
\DeclareUnicodeCharacter{391}{A}
\DeclareUnicodeCharacter{392}{B}
\DeclareUnicodeCharacter{395}{E}
\DeclareUnicodeCharacter{396}{Z}
\DeclareUnicodeCharacter{397}{H}
\DeclareUnicodeCharacter{399}{I}
\DeclareUnicodeCharacter{39A}{K}
\DeclareUnicodeCharacter{39C}{M}
\DeclareUnicodeCharacter{39D}{N}
\DeclareUnicodeCharacter{39F}{O}
\DeclareUnicodeCharacter{3A1}{P}
\DeclareUnicodeCharacter{3A4}{T}
\DeclareUnicodeCharacter{3A7}{X}

\DeclareUnicodeCharacter{27E6}{\relax\ifmmode \llbracket \else $\llbracket$\fi}
\DeclareUnicodeCharacter{27E7}{\relax\ifmmode \rrbracket \else $\rrbracket$\fi}
\DeclareUnicodeCharacter{00B5}{\relax\ifmmode \mu{} \else $\mu$\fi}
\DeclareUnicodeCharacter{1D434}{\relax\ifmmode A \else $A$\fi}
\DeclareUnicodeCharacter{1D435}{\relax\ifmmode B \else $B$\fi}
\DeclareUnicodeCharacter{1D436}{\relax\ifmmode C \else $C$\fi}
\DeclareUnicodeCharacter{1D437}{\relax\ifmmode D \else $D$\fi}
\DeclareUnicodeCharacter{1D438}{\relax\ifmmode E \else $E$\fi}
\DeclareUnicodeCharacter{1D439}{\relax\ifmmode F \else $F$\fi}
\DeclareUnicodeCharacter{1D43A}{\relax\ifmmode G \else $G$\fi}
\DeclareUnicodeCharacter{1D43B}{\relax\ifmmode H \else $H$\fi}
\DeclareUnicodeCharacter{1D43C}{\relax\ifmmode I \else $I$\fi}
\DeclareUnicodeCharacter{1D43D}{\relax\ifmmode J \else $J$\fi}
\DeclareUnicodeCharacter{1D43E}{\relax\ifmmode K \else $K$\fi}
\DeclareUnicodeCharacter{1D43F}{\relax\ifmmode L \else $L$\fi}
\DeclareUnicodeCharacter{1D440}{\relax\ifmmode M \else $M$\fi}
\DeclareUnicodeCharacter{1D441}{\relax\ifmmode N \else $N$\fi}
\DeclareUnicodeCharacter{1D442}{\relax\ifmmode O \else $O$\fi}
\DeclareUnicodeCharacter{1D443}{\relax\ifmmode P \else $P$\fi}
\DeclareUnicodeCharacter{1D444}{\relax\ifmmode Q \else $Q$\fi}
\DeclareUnicodeCharacter{1D445}{\relax\ifmmode R \else $R$\fi}
\DeclareUnicodeCharacter{1D446}{\relax\ifmmode S \else $S$\fi}
\DeclareUnicodeCharacter{1D447}{\relax\ifmmode T \else $T$\fi}
\DeclareUnicodeCharacter{1D448}{\relax\ifmmode U \else $U$\fi}
\DeclareUnicodeCharacter{1D449}{\relax\ifmmode V \else $V$\fi}
\DeclareUnicodeCharacter{1D44A}{\relax\ifmmode W \else $W$\fi}
\DeclareUnicodeCharacter{1D44B}{\relax\ifmmode X \else $X$\fi}
\DeclareUnicodeCharacter{1D44C}{\relax\ifmmode Y \else $Y$\fi}
\DeclareUnicodeCharacter{1D44D}{\relax\ifmmode Z \else $Z$\fi}
\DeclareUnicodeCharacter{1D44E}{\relax\ifmmode a \else $a$\fi}
\DeclareUnicodeCharacter{1D44F}{\relax\ifmmode b \else $b$\fi}
\DeclareUnicodeCharacter{1D450}{\relax\ifmmode c \else $c$\fi}
\DeclareUnicodeCharacter{1D451}{\relax\ifmmode d \else $d$\fi}
\DeclareUnicodeCharacter{1D452}{\relax\ifmmode e \else $e$\fi}
\DeclareUnicodeCharacter{1D453}{\relax\ifmmode f \else $f$\fi}
\DeclareUnicodeCharacter{1D454}{\relax\ifmmode g \else $g$\fi}
\DeclareUnicodeCharacter{1D456}{\relax\ifmmode i \else $i$\fi}
\DeclareUnicodeCharacter{1D457}{\relax\ifmmode j \else $j$\fi}
\DeclareUnicodeCharacter{1D458}{\relax\ifmmode k \else $k$\fi}
\DeclareUnicodeCharacter{1D459}{\relax\ifmmode l \else $l$\fi}
\DeclareUnicodeCharacter{1D45A}{\relax\ifmmode m \else $m$\fi}
\DeclareUnicodeCharacter{1D45B}{\relax\ifmmode n \else $n$\fi}
\DeclareUnicodeCharacter{1D45C}{\relax\ifmmode o \else $o$\fi}
\DeclareUnicodeCharacter{1D45D}{\relax\ifmmode p \else $p$\fi}
\DeclareUnicodeCharacter{1D45E}{\relax\ifmmode q \else $q$\fi}
\DeclareUnicodeCharacter{1D45F}{\relax\ifmmode r \else $r$\fi}
\DeclareUnicodeCharacter{1D460}{\relax\ifmmode s \else $s$\fi}
\DeclareUnicodeCharacter{1D461}{\relax\ifmmode t \else $t$\fi}
\DeclareUnicodeCharacter{1D462}{\relax\ifmmode u \else $u$\fi}
\DeclareUnicodeCharacter{1D463}{\relax\ifmmode v \else $v$\fi}
\DeclareUnicodeCharacter{1D464}{\relax\ifmmode w \else $w$\fi}
\DeclareUnicodeCharacter{1D465}{\relax\ifmmode x \else $x$\fi}
\DeclareUnicodeCharacter{1D466}{\relax\ifmmode y \else $y$\fi}
\DeclareUnicodeCharacter{1D467}{\relax\ifmmode z \else $z$\fi}

\DeclareUnicodeCharacter{1E67}{\.{\v s}}
\DeclareUnicodeCharacter{1E11}{\relax\ifmmode \c{d} \else $\c{d}$\fi}
\DeclareUnicodeCharacter{1ECB}{\relax\ifmmode \d{i} \else $\d{i}$\fi}
\DeclareUnicodeCharacter{1D8D}{\relax\ifmmode \textlhookx \else $\textlhookx$\fi}
\DeclareUnicodeCharacter{104}{\relax\ifmmode \k{A} \else $\k{A}$\fi}
\DeclareUnicodeCharacter{211E}{\relax\ifmmode \textrecipe \else $\textrecipe$\fi}
\DeclareUnicodeCharacter{29D}{\relax\ifmmode \textctj \else $\textctj$\fi}

\DeclareUnicodeCharacter{1E2E}{\'{\"I}}
\DeclareUnicodeCharacter{23F}{\textrts}

\DeclareUnicodeCharacter{2C73}{\varw}

\DeclareUnicodeCharacter{2127}{\mho}

\DeclareUnicodeCharacter{28C}{\textturnv}
\DeclareUnicodeCharacter{252}{\textturnscripta}
\DeclareUnicodeCharacter{259}{\schwa}
\DeclareUnicodeCharacter{25B}{\m{e}}
\DeclareUnicodeCharacter{266}{\m{h}}
\DeclareUnicodeCharacter{127}{\B{h}}
\DeclareUnicodeCharacter{27E}{\textfishhookr}
\DeclareUnicodeCharacter{281}{\textinvscr}

\continuouspages{}
\firstpage{1} 
\pubvolume{0}
\issuenum{0}
\elocationid{0}
\articlenumber{84915}
\pubyear{2026}
\copyrightyear{2026}
\datereceived{01 May 2026}
\dateaccepted{17 August 2026}
\dateonlinefirst{}
\datepublished{}

\usepackage[OT1,OT2,T2A,T2B,T2C,T3,T5,T1]{fontenc}
\usepackage[russian,english]{babel}

\usepackage{algorithm}
\usepackage{algorithmic}
\usepackage{color} 
\usepackage{soul} 
\usepackage[dvipsnames]{xcolor}
\usepackage{soul} 

\usepackage{tabularx}
\usepackage{lipsum}
\usepackage{tabularx,booktabs}
 \usepackage{multirow}
 \usepackage{graphicx}
 \usepackage[table,xcdraw]{xcolor}
\usepackage{colortbl}  
\usepackage[table,xcdraw]{xcolor}
\usepackage{color}
\usepackage{verbatim}
\usepackage{dblfloatfix}
\usepackage{tikz}
\usepackage{lineno}
\AtBeginDocument{%
  \nolinenumbers
  \renewcommand{\linenumbers}{}

}

\usepackage{soul}

{\color{red}{}}

\Title{Federated Cybersecurity Testbed as a Service (FCTaaS): A Framework to Federate Cybersecurity Testbeds}

\Author{
Josh Dean\textsuperscript{1}, 
Yu-Zheng Lin\textsuperscript{1}, 
John Paul Martin Encinas\textsuperscript{1},
Ibrahim Almazyad\textsuperscript{1}, 
Qinxuan Shi\textsuperscript{2}, 
Zhanglong Yang\textsuperscript{2}, 
Shalaka Satam\textsuperscript{1}, 
Tingjun Lei\textsuperscript{2}, 
Jielun Zhang\textsuperscript{2}, 
Sicong Shao\textsuperscript{2,}*, 
Salim Hariri\textsuperscript{1}, 
and Pratik Satam\textsuperscript{1,3,}*
}

\AuthorNames{
Josh Dean, 
Yu-Zheng Lin, 
John Paul Martin Encinas,
Ibrahim Almazyad, 
Qinxuan Shi, 
Zhanglong Yang, 
Shalaka Satam, 
Tingjun Lei, 
Jielun Zhang, 
Sicong Shao, 
Salim Hariri, 
and Pratik Satam}

\address{%
\textsuperscript{1}Department of Electrical and Computer Engineering, University of Arizona, Tucson, AZ, USA

\textsuperscript{2}School of Electrical Engineering and Computer Science, University of North Dakota, Grand Forks, ND, USA%

\textsuperscript{3}Department of Systems and Industrial Engineering, University of Arizona, Tucson, AZ, USA}
\corres{Corresponding Authors: Sicong Shao. Email:  sicong.shao@und.edu; Pratik Satam. Email: pratiksatam@arizona.edu}

\abstract{
Rapid technological change is transforming our society with the emergence of new fields such as Autonomous Vehicles and Smart Manufacturing, posing new research questions and challenges in system design, operations, security, and training. Researchers rely on testbeds to create experimental scenarios to solve these research challenges. These testbeds aid in data collection, analysis, and observation of the research problem, and in measuring the efficacy of the proposed solution. However, the rapid pace of modern innovation makes it challenging and cost-prohibitive for these testbeds to represent state-of-the-art scenarios, which require expensive upgrades and the addition of new capabilities. In addition, a small, select community of researchers has access to these specialized testbeds, which are not used optimally during their short operational lifecycles. This paper presents FCTaaS: Federated Cybersecurity Testbed as a Service, a scalable framework that bridges the current gaps in existing federation frameworks by enabling two heterogeneous cybersecurity testbeds to participate in a single experiment, overcoming testbed geographical boundaries to address today's research challenges. This federation allows multiple geographically isolated testbeds to work together to conduct experiments over a Virtual Private Network (VPN), while providing researchers with interfaces for testbed discovery and remote experiment design. Our experimental evaluation stresses the FCTaaS by assessing its performance in Denial-of-Service Scenarios on smart infrastructure and by evaluating intrusion detection and prevention workflows using a Suricata-based intrusion detection and intrusion prevention system (IDS/IPS) testbed. The limiting network utilization of 49\% and low overhead of 1\%, even under such resource-intensive attack scenarios, demonstrates the efficacy of FCTaaS across three case studies thus enabling improved experimentation while preserving visibility into attack traffic, IDS alerts, and detection-system resource stress with latency ranging from 5.63 ms between local nodes to 147 ms between disbursed nodes.
}


\begin{document}

\section{Introduction} \label{sect:s1}
The growing integration of communication networks, artificial intelligence (AI), and automated control has expanded the scale and complexity of cyber-physical systems (CPS). These developments are central to applications such as smart manufacturing, autonomous systems, energy infrastructure, and digital twins. At the same time, they introduce new challenges in system design, operational management, cybersecurity, and workforce training. In particular, the adoption of Industry 4.0 technologies requires security evaluation methodologies that can capture interactions among physical processes, control interfaces, networked services, and adversarial behaviors.

Specialized testbeds provide an essential foundation for such evaluation. They enable researchers to reproduce representative operational conditions, collect data, assess security mechanisms, and study the effects of cyberattacks on physical processes in controlled environments. However, CPS-security testbeds are often expensive to build and maintain, require dedicated space and technical personnel, and are typically developed to address specific research objectives. Consequently, individual testbeds may remain underutilized or may lack complementary capabilities required for more comprehensive experiments. For example, a physical CPS testbed may need to interact with a cloud-hosted supervisory control service, an attacker environment, an intrusion detection and prevention system (IDS/IPS), and a monitoring platform. Reproducing all such capabilities at a single site can be costly and difficult to sustain as technologies and security requirements evolve.

Federating geographically distributed testbeds offers a practical approach for combining complementary capabilities without requiring physical co-location. Nevertheless, CPS-security federation introduces requirements beyond general-purpose networking or cloud experimentation. Participating testbeds may be independently managed, expose different control and communication interfaces, enforce distinct access and availability policies, and serve different experimental roles. A usable federation framework must therefore support policy-aware testbed onboarding, experiment lifecycle coordination, controlled connectivity, and integrated observation across attacker, CPS, control, IDS/IPS, and monitoring~components.

This paper presents the \emph{Federated Cybersecurity Testbed as a Service} (FCTaaS) framework, which enables independently managed and geographically distributed CPS-security testbeds to participate in a common experiment. In this work, federation refers to the coordinated composition of semi-autonomous testbeds into a shared experimental workflow while preserving testbed-owner control over access and availability. FCTaaS combines Testbed Managers, REST-based interfaces, access-policy configurations, Kafka-based coordination, and experiment-specific virtual private network (VPN) connectivity. These components support testbed discovery, authorization, initialization, experiment execution, monitoring integration, and teardown. Existing federation platforms have made important contributions to networking, cloud, and general-purpose experimentation. However, CPS-security experiments often require additional support to coordinate physical assets, industrial control services, attacker environments, IDS/IPS components, and monitoring services within a single workflow. FCTaaS addresses this need by providing a CPS-security-oriented federation and orchestration layer that supports the coordinated composition of heterogeneous testbeds into a common experimental workflow.

The principal contributions of this work are as follows:
\begin{itemize}
    \item We introduce FCTaaS, a federation and experiment-orchestration framework for integrating geographically distributed and independently managed CPS-security testbeds into common experimental workflows.

    \item We provide policy-aware testbed onboarding and access control through Testbed Description Files, Access Policy Files, Testbed Managers, and experiment lifecycle management services, allowing testbed owners to retain control over resource availability and authorized usage.

    \item We support the composition of heterogeneous experimental roles, including physical CPS assets, cloud-hosted control services, attacker environments, IDS/IPS components, and monitoring platforms, with coordinated initialization, secure experiment-specific connectivity, and teardown.

    \item We demonstrate the practical feasibility of FCTaaS through federated CPS-security case studies involving Denial-of-Service (DoS), IDS/IPS detection and prevention, resource monitoring, and Man-in-the-Middle (MITM) attack workflows.

    \item We characterize deployment-oriented performance observations, including end-to-end communication delay, resource behavior under attack-driven IDS/IPS workloads, and the incremental host-resource cost associated with integrating the evaluated Smart Water Testbed into FCTaaS.
\end{itemize}
	
The remainder of the paper follows the following structure. In Section II, we present the related work where we discuss cybersecurity testbeds. Section III presents the FCTaaS approach and architecture and the required components for successfully integrating the multiple heterogeneous cybersecurity testbeds into a single federation. In Section IV, we present experimental results from demonstrated experiments and discuss them. Next, section V quantifies the overhead of the federation framework.  Finally, section VI finalizes the paper's conclusion.


\section{Related Work} \label{sect:s2}

This section highlights the related work on the Federated Cybersecurity Testbed as a Service (FCTaaS) architecture presented in this paper.

\subsection{Testbed Federation} \label{sect:s2dot1}
The federation model relies on multiple resources owned and operated by different providers functioning together to provide a service. In the context of the FCTaaS, the federation model focuses on providing interoperability and communications between diverse testbeds that are by default, not interoperable. Such a federation allows researchers to design and perform experiments not possible with each of the testbeds individually. 

\begin{itemize}
    \item \textbf{Emulab:} Emulab \cite{Hibler2008Large-scaleTestbed,Garcia2010}, built-in 2008, is the first attempt at testbed federation over the Internet. Emulab provides an infrastructure to emulate complex networking environments for networking experimentation. Built on the assumption that pure software simulation is too simplistic for complex networking environments, the Emulab architecture consists of two control servers called the boss and ops. The boss and ops help control and manage the pool of interconnected nodes available in the computation cluster for Emulab experimentation. The boss server provides a web interface, while the ops server enables access to the experimental nodes. Experiments in the Emulab environment are configured through NS language, allowing setup and evaluation of complex networking topologies.
    \item \textbf{GENI:} Global Environment for Networking Innovation (GENI), extends Emulab to bridge geographically dispersed sites to allow federation of resources. GENI (funded by the National Science Foundation (NSF)) is the first large-scale effort to enable the federation of heterogeneous testbeds \cite{Berman2014GENI:Experiments}. GENI enables experimentation on computer network security, software-defined networking, distributed computing, and cloud computing \cite{Goodfellow2018Merge:Ecosystems,Baldin2018TheInfrastructure}. However, GENI has several shortcomings, such as, (1) The researchers do not have a standardized framework to integrate new systems into the federation; (2) Unable to access real-world systems due to network traffic and Quality of Service (QoS) limitations; (3) High initial setup costs as a GENI site requires dedicated hardware to run a GENI node \cite{Baldin2018TheInfrastructure,Friedman2019TheSystem,Bavier2016Planetignite:Cloud}. The proposed FCTaaS overcomes the shortfalls of GENI by providing a unified way to integrate new devices while keeping a low integration cost and an easy integration process.
    \item \textbf{PlanetLab:} PlanetLab \cite{peterson2006planetlab} is a global research network that uses Slice-Based Federation Architecture (SFA) \cite{onelab_sfa}, allowing academic research organizations to share their computational and networking resources for better experimentation. The PlanetLab creates an overlay network on top of the Internet, allowing experiment support over layer three for experimentation. PlanetLab and its SFA, are the reference points for other successful testbeds, including the GENI.
    \item \textbf{CloudLab:} CloudLab\cite{duplyakin2019design} is a software technology designed to empower researchers in creating and overseeing cloud computing environments. It is built upon the foundation of Emulab \cite{White+:osdi02} and various GENI software technologies. CloudLab provides users with direct access to computing, network, and storage resources. This platform enables experimenters to execute cloud software stacks, accommodating a spectrum of experiments, ranging from brief targeted tests to enduring, production-quality cloud setups. The interconnected CloudLab sites establish communication through both IP and Layer-2 links, connecting to regional and national research networks. The clusters, situated in Utah, Wisconsin, Clemson, and the GENI testbed, collectively contribute to the capabilities of CloudLab \cite{University_of_Utah_2022}.

    \item \textbf {Fed4Fire+:} The European Facility for Future Internet Research and Experimentation focuses on specific communities within the realm of the Future Internet \cite{Fed4FIRE,vandenberghe2013architecture}. Its research infrastructures enable experiments that span different domains, facilitated by Fed4FIRE+ through a unified federation framework. The federation brings together existing experimentation facilities in Europe that specialize in networking-related research or cater to varied service and application communities, including communication platform, Internet of Things platform, and cloud platform~\cite{yala2019testbed,gaglianese2022lightweight}. Moreover, the fundamental limitations of this framework are: (1) Focus on orchestrating homogeneous testbeds for experimenting with research applications, and it does not integrate unmatched communication protocols; (2) The platform lacks support for cybersecurity tools for conducting experiments and evaluating testbed performance under the cyber attack. (3) After assigning a testbed to users, it does not have oversight and monitoring. It forces additional efforts to improve the framework's capabilities. 
\end{itemize}


Table \mbox{\ref{lit_tab:ad}} provides a feature-oriented comparison of representative federation platforms based on the capabilities most relevant to the scope of this work. The comparison is not intended to imply that the referenced platforms cannot be extended with additional cybersecurity tools or domain-specific services. Rather, it highlights the design focus of FCTaaS: the coordinated integration of independently managed CPS-security testbeds, including policy-aware access control, testbed lifecycle management, and the composition of attacker, CPS, Supervisory Control and Data Acquisition (SCADA), IDS/IPS, and monitoring components within a common experimental workflow. In this comparison, cross-protocol experimentation refers to the ability to orchestrate experiments involving testbeds that expose different communication and control interfaces; it does not imply universal protocol translation or semantic mediation. Similarly, the platform performance evaluation capability refers to the structured set of evaluation dimensions used in this work, rather than to a deployment-independent performance benchmark. FCTaaS is therefore positioned as a CPS-security-oriented federation and orchestration framework, whereas the cited platforms primarily emphasize networking, cloud, or general-purpose experimentation resources.

\subsection{CPS-Security Testbed Federation Requirements}

Cyber-physical system (CPS) security experimentation requires test environments that can represent interactions among physical processes, control logic, communication networks, security mechanisms, and human or organizational constraints. The NIST CPS Framework identifies trustworthiness, timing, data, boundaries, composability, and lifecycle as cross-cutting CPS concerns \cite{cyber2017framework}. These concerns become particularly important when an experiment spans multiple independently managed testbeds \cite{baumgartner2010virtualising,coulson2012flexible}, because the resulting workflow must coordinate resources that differ in location, ownership, operational role, communication interface, and security policy. Prior work on cybersecurity experimentation emphasizes the need for environments that support controlled experimental execution, meaningful observation, credible system representation, and repeatable experimentation \cite{tunc2015claas,benzel2011science}. Similarly, CPS testing research highlights the tradeoffs among realism, scalability, cost, and reproducibility when selecting physical, virtual, simulated, or hybrid test environments \cite{zhou2018review,chen2014implementing}. These considerations imply that a federated CPS-security platform should be evaluated not only according to whether participating testbeds can be connected, but also according to whether the resulting environment enables researchers to operate, observe, reproduce, and meaningfully represent the intended experimental scenario.

\begin{table}[H]
\caption{A comparative examination of existing federation architectures in relation to the proposed FCTaaS framework, based on capabilities most relevant to cybersecurity experimentation on cyber-physical systems. The comparison highlights that the design focus of FCTaaS is on coordinated integration of independently managed CPS-security testbeds, together with support for cyber assessment, performance evaluation, user-driven testbed integration, and cross-protocol experimentation.}
\label{lit_tab:ad}
\resizebox{\textwidth}{!}{%
\begin{tabular}{lccccc}
\hline
\textbf{\begin{tabular}[c]{@{}l@{}}Federation\\ Architecture\end{tabular}} & \textbf{\begin{tabular}[c]{@{}c@{}}Specialized Cyber-Physical\\ Testbeds Integration\end{tabular}} & \textbf{\begin{tabular}[c]{@{}c@{}}Facilitate Cyber \\ Assessment Tools\end{tabular}} & \textbf{\begin{tabular}[c]{@{}c@{}}Platform Performance\\ Evaluation Assessment\end{tabular}} & \textbf{\begin{tabular}[c]{@{}c@{}}Allow Users\\ Testbeds Integration\end{tabular}} & \textbf{\begin{tabular}[c]{@{}c@{}}Enable Cross-Protocol\\ Experimentation\end{tabular}} \\ \hline
Emulab        & Emulation only                   & Add-on by user     & Not Provided     & Not Supported      & Not Supported  \\
PlanetLab     & Emulation only                   & Add-on by user     & Not Provided     & Not Supported      & Not Supported  \\
GENI          & Emulation only                   & Add-on by user     & Not Provided     & Not Supported      & Not Supported  \\
CloudLab      & Emulation and Cyber-Physical     & Add-on by user     & Not Provided     & Limited Option     & Not Supported  \\
Fed4Fire+     & Emulation and Cyber-Physical     & Add-on by user     & Not Provided     & Limited Option     & Not Supported  \\
\hline
FCTaaS        & Cyber-Physical                   & Platform Native    &  Provided        & Open For Users     & Supported  \\
\hline
\end{tabular}%
}
\end{table}

Federation introduces additional requirements beyond those of a single-site testbed \cite{ricci2012designing}. Participating resources may expose different control and monitoring interfaces, enforce different access and availability constraints, and operate under different timing and network conditions \cite{agarwal2016unified}. CPS federation environments such as the Universal CPS Environment for Federation demonstrate the importance of composing heterogeneous tools and federates into consolidated experiments \cite{song2019ieee,burns2018universal,roth2017cyber}. In addition, timing calibration studies of CPS testbeds show that communication delay, synchronization uncertainty, and deployment conditions can materially affect the interpretation of experimental results \cite{weiss2018timing}. Therefore, a federation framework must preserve testbed-owner control while coordinating experiment setup, connectivity, monitoring, and teardown across distributed participants.

These requirements motivate the four evaluation dimensions used in this work. \emph{Controllability} characterizes the extent to which authorized researchers can configure and operate a participating testbed through the federation. \emph{Observability} characterizes the visibility available into cyber events, control communication, physical-system behavior, and resource conditions during an experiment. \emph{Repeatability} concerns the ability to preserve and reinstantiate the relevant testbed selection, configuration, initialization, and experimental conditions. \emph{Fidelity} concerns the extent to which the federated workflow preserves the salient physical, control, communication, and security interactions required by the intended CPS-security scenario. These dimensions are complementary. Increasing federation scale and remote accessibility can reduce direct physical control, introduce communication delay, or limit access to local instrumentation. Conversely, tighter control and higher-fidelity physical integration can increase operational cost and reduce the ease with which experiments are shared or replicated across sites. Section~\ref{sec:exp_metrics} operationalizes these dimensions as a structured evaluation framework for the FCTaaS case studies.

\section{FCTaaS Architecture} \label{sect:s3}

\begin{figure}[H]
    \centering
        \includegraphics[width=.8\columnwidth]{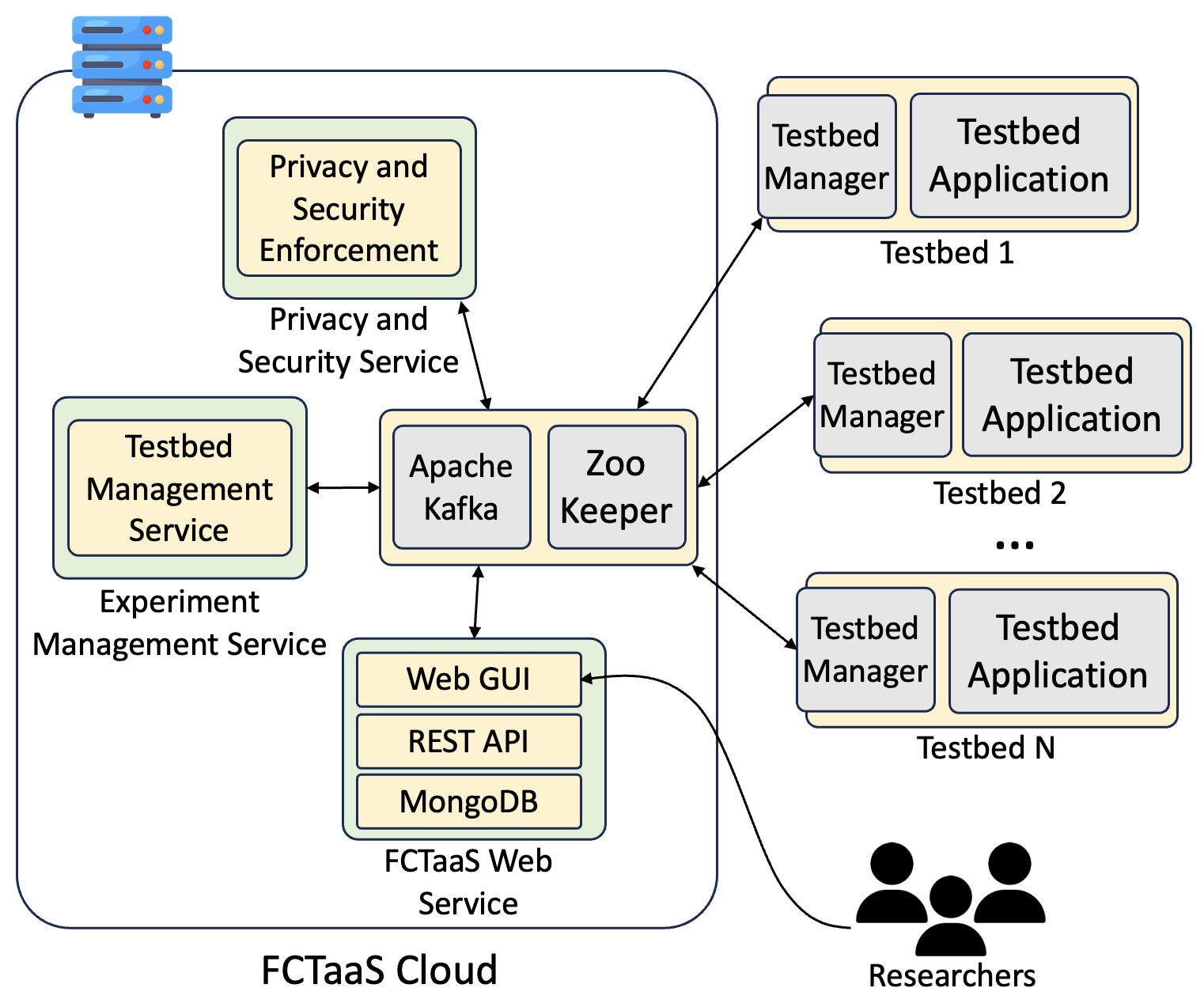}
        \caption{FCTaaS Architecture. The figure illustrates how the FCTaaS Web Service, Experiment Management Service, Privacy and Security Service, and Kafka-based interoperability layer coordinate multiple independently managed testbeds to enable a single federated experiment.}
  \label{fig:Architecture}
\end{figure}

FCTaaS is designed to coordinate independently managed cybersecurity testbeds that differ in location, operational role, access policy, and communication interface. The architecture follows four design principles. First, \emph{incremental scalability} allows new testbeds to join through a Testbed Manager, a testbed description, an access-policy configuration, and initialization logic. Second, \emph{testbed-level interoperability} enables heterogeneous CPS, control, attacker, IDS/IPS, and monitoring components to participate in a common experimental workflow. Third, \emph{policy-aware security} preserves testbed-owner control through authorization checks, access policies, and experiment-specific connectivity. Fourth, \emph{latency-aware deployment} recognizes that distributed experiments remain subject to wide-area network conditions, endpoint resources, and traffic-processing overhead. Fig.~\ref{fig:Architecture} illustrates the architecture. The Interoperability Service coordinates management-plane communication among FCTaaS services and Testbed Managers. The Experiment Management Service coordinates testbed selection, reservation, initialization, and teardown. The Privacy and Security Service verifies authorization according to testbed-owner-defined policies. The FCTaaS Web Service provides interfaces for testbed discovery, experiment configuration, and experiment execution. During an approved experiment, the selected testbeds are connected through an experiment-specific VPN overlay, while application and attack traffic remain on the configured experimental data path. Section~\ref{sec:exp_metrics} subsequently evaluates the operational tradeoffs of this design using controllability, observability, repeatability, and fidelity.

Fig. \mbox{\ref{fig:Architecture}} illustrates that the proposed architecture separates experiment coordination from the internal implementation of each participating testbed. This separation is important because it allows each testbed to join the federation through its Testbed Manager without requiring direct custom integration with every other testbed. As a result, the architecture improves scalability, preserves testbed autonomy, and enables the coordinated use of geographically distributed cybersecurity testbeds.

\subsection{Individual Testbeds} \label{sect:s3dot1}

Cybersecurity testbeds are designed and developed for experimenting on specific research problems and are expensive to develop and maintain \cite{Craggs2019ATestbeds}. Nevertheless, these testbeds are often underutilized and inactive when used by an individual organization \cite{KeaheyLessonsTestbed}. Testbed Manager (TM) API enables the integration of individual testbeds into the FCTaaS federation environment. Using the TM, the testbed owners integrate the testbed into the FCTaaS, allowing researchers to use the testbed remotely while utilizing the testbed in a larger scale experiment. Using the FCTaaS Web Interface or API calls, testbed owners can register their testbed by providing a Testbed Description File and Access Policy File. The Testbed Description file describes the testbed capabilities and utilization, while the Access Policy file describes the testbed access conditions and which users can use the testbed. Through the Testbed Description File and Access Policy File, the testbed owner easily integrates their unique hardware into the federation while enforcing their access control policies. 

 A new testbed while joining the federation goes through a registration phase, where the testbed manager interacts with the interoperability service to identify the testbed uniquely with an identifier, and share the testbed capabilities defined in the Description File with the federation. After the registration phase, FCTaaS lists this testbed in the FCTaaS testbed repository allowing its usage for federated experimentation. Researchers can browse available testbeds in the FCTaaS testbed repository, observe its description and access policies, and select necessary testbeds. Researchers then initialize testbed parameters, configure network devices, and establish the desired configuration. Once the experiment is over, the user can finalize the experiment through the web interface or API calls, and FCTaaS will decompose all of the testbeds involved. Researchers can efficiently reproduce this experiment by saving the configuration parameters and testbeds involved in the experiment in their user profiles.

\subsection{FCTaaS Web Service} \label{sect:s3dot2}
The FCTaaS Web Service provides the primary user-facing interface for accessing and managing the federation. It allows researchers to: (1) discover testbeds registered in the federation, (2) examine their capabilities and access conditions, (3) configure federated experiments, and (4) initiate, monitor, and terminate experiment execution. The Web Service consists of three main components: a MongoDB database for persistent storage, a REST-based application programming interface for management-plane communication and experiment lifecycle management, and a web-based graphical user interface through which researchers interact with the FCTaaS environment. The MongoDB database stores user profiles, testbed descriptions, access-policy information, testbed availability and operational status, experiment configurations, and experiment records. This persistent storage enables researchers to retain the testbeds and configuration parameters associated with an experiment, thereby supporting subsequent experiment reconstruction and execution. Testbed owners register their resources by providing a Testbed Description File and an Access Policy File, which describe the capabilities, initialization requirements, availability constraints, and permitted user groups associated with each testbed.

The FCTaaS REST interface supports synchronous management-plane communication among the Web Service, the Experiment Management Service, the Privacy and Security Service, and the Testbed Managers associated with participating testbeds. At the functional level, the interface supports testbed registration, testbed discovery and metadata retrieval, user authentication and authorization, testbed-availability checking, experiment submission, testbed initialization, operational-status reporting, and experiment termination. Through these operations, Testbed Managers exchange testbed descriptions, access-policy information, experiment parameters, initialization results, availability information, and operational status with the FCTaaS services. The same interface provides the backend functions required by the Web GUI, including operations for registering a new testbed, listing available testbeds, retrieving testbed information, submitting an experiment configuration, checking experiment status, and stopping an active experiment. The REST interface is used for experiment coordination and lifecycle management rather than for transporting the application or attack traffic generated during an experiment. After an experiment request is submitted, the Experiment Management Service coordinates authorization, testbed reservation, initialization, and teardown through the corresponding Testbed Managers. Asynchronous lifecycle events and testbed-state updates are exchanged through the Kafka-based publish-subscribe Interoperability Service described in Section 3.5. This separation prevents the Web Service from requiring continuous synchronous communication with every participating testbed and enables state changes to be disseminated to the relevant FCTaaS services.

Once the selected testbeds have been authorized and initialized, they are connected through an experiment-specific VPN overlay. Experimental traffic is then exchanged directly among the participating testbeds through the VPN using the protocols required by the individual testbed applications. For example, in the Smart Water case study, the ScadaBR Testbed communicates with the Smart Water Testbed using Modbus/TCP, while attack traffic generated by the Virtual Cybersecurity Attacker Testbed is also transported through the experiment VPN. Therefore, the FCTaaS communication architecture separates REST-based synchronous lifecycle management, Kafka-based asynchronous coordination and state dissemination, and VPN-based experimental data communication using testbed-specific application protocols. This separation allows independently managed and heterogeneous testbeds to participate in a common experiment without requiring their internal applications to adopt a single communication protocol.

\subsection{Experiment Management Services} \label{sect:s3dot3}

The Experiment Management Service (EMS) \(S_{mg}\) coordinates the individual Testbed Managers to orchestrate federated experiments. Algorithm \ref{alg:one} summarizes the nominal experiment-creation workflow in FCTaaS when the requested testbeds are authorized, available, and successfully initialized. In Algorithm \ref{alg:one}, the FCTaaS Cloud \(C_{fct}\) denotes the logical coordination environment through which the FCTaaS Web Service and the supporting management services are exposed to the researcher. It does not represent an additional service separate from the architectural components illustrated in Fig. \ref{fig:Architecture}. Similarly, an interaction with a testbed \(tb\) in the algorithm is mediated by the Testbed Manager associated with that testbed rather than being sent directly to the underlying physical or virtual resource. An experiment begins when a researcher submits a single experiment-creation request through the FCTaaS Web Service. The FCTaaS Cloud forwards the request to the Privacy and Security Service \(S_{ps}\), which verifies whether the researcher is authorized to access the requested testbeds according to the applicable Access Policy Files. After authorization is granted, the FCTaaS Cloud requests the Experiment Management Service to determine the availability of the required testbeds. The Experiment Management Service then obtains the availability status of each requested testbed through its associated Testbed Manager.

The condition ``Requested testbeds are available'' in Algorithm \mbox{\ref{alg:one}} represents an aggregate condition over the complete set of requested testbeds. Availability is therefore confirmed to the FCTaaS Cloud only after all requested testbeds have returned an available status. The individual availability requests are shown within the iterative structure to represent the collection of status information from each participating testbed. Once authorization and aggregate availability have been confirmed, the Experiment Management Service proceeds with experiment initialization. At the level of abstraction used in Algorithm \ref{alg:one}, the initialization request includes reservation of the selected testbeds for the requested experiment period, distribution of the experiment-specific configuration parameters, and provisioning of the network information required to join the experiment VPN. Each Testbed Manager executes the applicable local initialization procedure, configures its associated testbed for the experiment, and returns an initialization-completion status to the Experiment Management Service. The condition ``All testbeds finished initialization'' represents a synchronization point in the experiment-creation workflow. The Experiment Management Service completes experiment creation only after every selected testbed has reported successful initialization. The experiment identifier and applicable credentials are then returned to the FCTaaS Cloud and presented to the authorized researcher through the FCTaaS Web Service.

\begin{algorithm}[H]
 \caption{: FCTaaS experiment creation algorithm} \label{alg:one}
 \begin{algorithmic}[1]
 \renewcommand{\algorithmicrequire}{\textbf{Input:}}
 \renewcommand{\algorithmicensure}{\textbf{Output:}}
 \REQUIRE Experiment testbeds: TBs=$\{{tb}_1,{tb}_2, ...,{tb}_n\}$,\\
 FCTaaS Cloud $C_{fct}$,
 \\Privacy and Security service $S_{ps}$,
 \\Experiment Management Service $S_{mg}$.
 \ENSURE Experiment credentials $K_{ex}$ and access instruction $D_{ac}$.
 \\
 \
  \STATE User sends experiment creation request to $C_{fct}$ via FCTaaS Web Service.
  \STATE $C_{fct}$ forwards the request to $S_{ps}$ and waits for response.
  \STATE $S_{ps}$ grants the access to the user, and sends back the permission to $C_{fct}$.
  \STATE $C_{fct}$ sends availability request to $S_{mg}$.
  
  \FOR{\textbf{each} $tb \in TBs$}
    \STATE $S_{mg}$ sends availability request to $tb$ and waits for the response.
    \STATE $tb$ sends the availability status to $S_{mg}.$
    \IF {(Requested testbeds are available)}
        \STATE $S_{mg}$ confirms the availability to $C_{fct}.$
    \ENDIF
  \ENDFOR
  
  \FOR{\textbf{each} selected $tb \in TBs$}
    \STATE $S_{mg}$ sends initialization request to $tb$.
    \STATE $tb$ finishes initialization and returns the completed message to $S_{mg}$.
    \IF {(All testbeds finished initialization)}
        \STATE $S_{mg}$ sends the experiment id and credential $K_{ex}$ to $C_{fct}.$
    \ENDIF
  \ENDFOR
  
 \STATE \textbf{return} $K_{ex}$ and $D_{ac}$
 \end{algorithmic}
\end{algorithm}

FCTaaS establishes an experiment-specific Virtual Private Network (VPN) overlay among the selected testbeds to support federated experimentation. During initialization, the Experiment Management Service provides the relevant VPN configuration parameters to each Testbed Manager. Each Testbed Manager applies the received configuration and prepares its associated testbed to participate in the experiment. Once all selected testbeds have completed initialization and joined the VPN, the Experiment Management Service performs a status check to confirm that the federated environment is ready. The researcher can then conduct the experiment through the FCTaaS Web Service. The VPN provides secure connectivity among the participating testbeds, while the Experiment Management Service and the individual Testbed Managers remain responsible for authorization coordination, availability checking, initialization, status reporting, and teardown. Experimental application, control, and attack traffic is exchanged among the participating testbeds through the configured VPN and, when applicable, through inline security components such as the IDS/IPS testbed.

\subsection{Privacy and Security Service} \label{sect:s3dot4}
The Privacy and Security Service module manages the authentication and authorization of all the FCTaaS activities. The Privacy and Security Service interacts with the EMS, during an experiment initialization, verifying the access a user has to a testbed \cite{Coyne2013ABACManagement}. This access is verified by the Testbed Manager using the Access Policy File (APF) set up by the testbed owner when the testbed joins the federation. Each APF has two access policy attributes: user-group attribute and availability-time attribute. The user-group attribute defines the level of access each user-type has to the testbed, wherein the user-type testbed owner has the most access, while the user-type federated-researcher has a degraded access to the testbed. The attribute availability-time determines and controls the availability times of the testbed for the FCTaaS experiments. For example, the testbed owner can specify the availability time in the access policy file from 8 AM to 5 PM, resulting in the rejection of requests to use the testbed at 6 PM. However, the testbed owner may choose not to specify the available time, enabling researchers to utilize the testbed 24/7.

The VPN-based federation is the primary tradeoff between security and performance. The encrypted communication mitigates concerns about eavesdropping or data tampering. However, the imposed latency accounts for the bulk of measured delay times. VPN endpoints also introduce additional attack surfaces; these are mitigated by generating credentials during initialization and by the access policy file, which restricts access to authorized participants.

\subsection{Interoperability Service} \label{sect:s3dot5}
The Interoperability Service manages all the communication between the testbeds while connected to the FCTaaS. The Interoperability Service enables the FCTaaS Web Service, Experiment Management Service, and the Privacy \& Security Service to communicate with each other and the Testbed Managers, using Apache Kafka-based publish-subscribe communication \cite{ApacheKafka,Esposito2013InterconnectingService}. The Kafka has four main components: producer, consumer, broker, and zookeeper. The producer is an agent that publishes messages to the broker(s), while a broker is responsible for disseminating this message to consumers requesting the produced messages. The zookeeper manages a cluster of brokers, allowing the realization of communication messages through topics. When a publisher publishes messages for a topic, all the consumers subscribed to the topic will receive the message. Kafka was selected because it naturally isolates communications and supports stateful monitoring of testbeds.

\subsection{Testbed Manager} \label{sect:s3dot6}
The Testbed Manager interfaces each testbed to the FCTaaS using the FCTaaS REST API, by implementing functionalities that allow the EMS to create experiments and initialize the testbed into the experiment, while part of the proposed federation. The Testbed Manager also allows the testbed owner to control and define the researcher's access to the testbed over the federation. This access control is performed with the Access Policy File (APF). The testbed owner also defines the testbed description, an XML file containing the Testbed ID (a unique identifier identifying the testbed in the federation), testbed capability descriptors, and testbed initialization script.

\subsection{FCTaaS Features} \label{sect:s3dot7}

The proposed framework encompasses several key features:

\begin{enumerate}
\item 
Integration and authentication APIs: The framework incorporates RESTful APIs for web interface integration, user authentication and authorization, as well as testbeds integration and instantiation.
\item 
Distributed oversight with Kafka: The framework integrates Kafka, a distributed streaming platform that enables real-time monitoring of available testbeds, current operational status, and maintenance of usage statistics for oversight purposes.
\item 
VPN for experiment federation: Virtual Private Networks are employed by the framework to facilitate the federation of experiments across different testbeds while ensuring secure and reliable communication throughout the experimentation process.
\item 
Support for heterogeneous experimentation: The framework is designed to expand its capabilities by accommodating experiments involving various communication protocols and underlying hardware.
\item 
Evaluation of performance metrics for the federation platform: The framework includes features to assess the performance metrics of the federation platform including controllability, observability, repeatability and fidelity, ensuring continuous improvement and optimization of the system.
\item 
Incorporation of established cybersecurity tools alongside innovative ones: Prioritizing security enhancement through integrating cybersecurity tools allows assessment of potential vulnerabilities within testbeds using specialized software developed either internally at the Autonomic Computing lab or from readily available sources. In particular, FCTaaS enables intrusion detection-centered experiments by federating IDS/IPS tools such as Suricata with cyber-physical testbeds, attacker testbeds, and monitoring tools so researchers can evaluate both attack detection accuracy and IDS resource behavior under realistic traffic loads.

\end{enumerate}

\subsection{Experimentation Platform Performance Metrics} \label{sec:exp_metrics}
\begin{figure}[H]
    \centering
  \includegraphics[width=0.6\columnwidth]{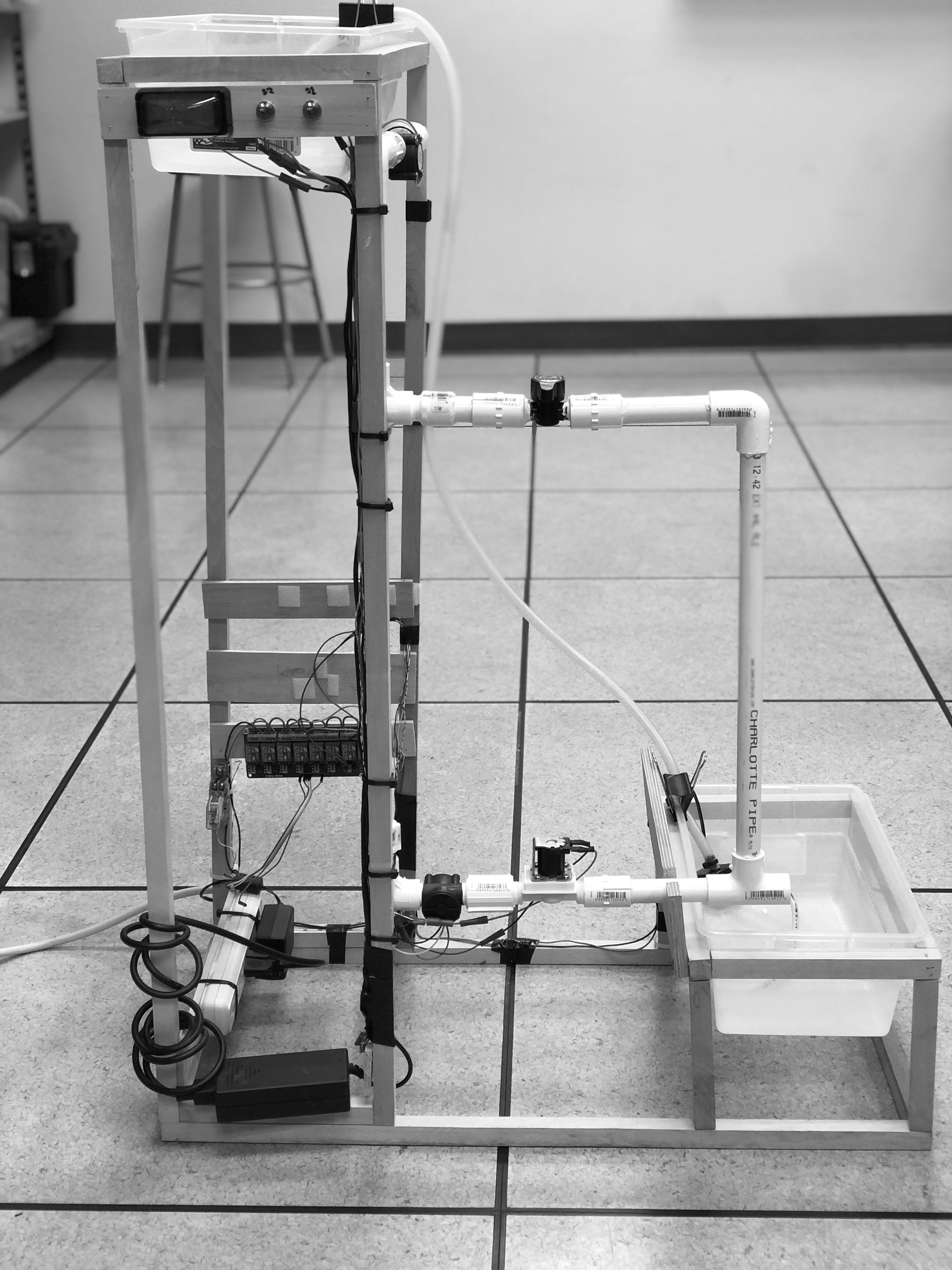}
  \caption{Smart Water Testbed. The figure shows the physical cyber-physical system used in the case studies, including the sensing, control, and actuation components that allow the impact of cyber-attacks to be observed at the operational level.}
  \label{fig:SW_TB}
\end{figure}

The evaluation framework adopts a multidimensional assessment strategy inspired by the DREAD threat-modeling \mbox{\cite{hussain2014threat}} approach, in which several interpretable dimensions are combined to support a structured overall assessment. Rather than characterizing federated experimentation through a single performance indicator, FCTaaS is evaluated across four complementary dimensions: controllability, observability, repeatability, and fidelity. These dimensions capture the extent to which a federated environment enables researchers to operate testbeds remotely, observe experimental behavior, reproduce experimental configurations, and preserve representative system interactions. The resulting ratings provide a common analytical basis for comparing the operational characteristics of the participating testbeds within each case study. They are assigned using evidence collected during experiment deployment and execution, including accessible control functions, available monitoring signals, configuration persistence, network conditions, and the degree to which the federated workflow represents the intended CPS-security scenario. The ratings are interpreted together with the quantitative measurements reported in the case studies, including attack-success behavior, communication latency, IDS/IPS resource utilization, and federation-related host-resource overhead.
\begin{itemize}
    \item \textbf{Controllability:} Controllability measures the degree of control and modifiability researchers have on a testbed. For instance, an in-person researcher using the Smart Water Testbed (Fig. \ref{fig:SW_TB}) has the highest controllability as they can modify and use the testbed as per their experimentation needs, including rewiring the valves and sensors, reprogramming the PLCs, and changing the network configurations. \\Meanwhile, using such testbeds through infrastructures similar to the proposed FCTaaS, causes a loss of controllability, restricting the changes a researcher can make to the testbed. 
    This controllability metric aims to quantify these restrictions imposed on a researcher due to the federation, comparing the control possible with a similar in-person testbed. This work measures the controllability metric as a ratio of access or capabilities that a researcher has to a testbed through the proposed FCTaaS to in-person access. While performing this measurement, the work concedes the inability of the FCTaaS to support physical or functional modifications to testbeds (like rewiring sensors or changing network configurations) and does not account for these losses in the measurement.
    \\
    The controllability metric evaluates the federation platform's ability to manage testbeds based on the following criteria: for cyber-physical systems that operate mechanically and are not remotely accessible (Very Low); for systems that partially benefit from the federation by allowing limited control of the system (Low); for systems that moderately benefit from the federation by enabling partial control of the system (Medium); for systems that fully benefit from the federation and have comprehensive control over testbeds, with minimal imposed limitations (High); and for systems that fully benefit from the federation and achieve complete automation and control over testbeds (Very High).

    \item \textbf{Observability:} Observability measures the degree of accurate measurements a researcher can make on a testbed during experimentation. Similar to the controllability, observability is reduced by integrating the testbed into infrastructure similar to the proposed FCTaaS. This reduction in observability is attributed to reduced controllability, and overhead added due to the infrastructure. This work measures observability as an error metric comparing measurements made in the FCTaaS vs the in-person measurements, with a lower value in error metric signifying higher observability.
    \\
    Observability provides a method for assessing the federation oversight of testbeds within the federation platform, with reference to the following criteria: For cyber-physical systems operating in a conservative manner and allowing very limited access to testbed measurements (Very Low); For systems benefitting selectively from the federation by enabling limited access to measurements of the system (Low); For cyber-physical systems that operate semi-privately and allow selective access to testbed measurements (Medium); For systems fully benefiting from the federation and establishing comprehensive oversight over testbeds, although there are few restrictions imposed (High); and For cyber-physical systems operating with complete accessibility and offering full oversight of testbed measurements (Very High).
    
    \item \textbf{Repeatability:} The Repeatability metric assesses the ability of the federation platform to replicate experiments multiple times based on the following criteria: The platform restricts experiment replication or fails to maintain experiment parameters (Low); The platform permits experiment replication but does not ensure maintain experiment parameters or guarantee consistent results (Medium); The platform enables experiment replication, maintains experiment parameters, and ensures consistent results~(High).
    
    \item \textbf{Fidelity:}
    The Fidelity metric provided assessment for the federation platform to accurately represent real system per the following criteria: for systems operating in isolated environments without leveraging the federation or interacting with other systems (Low); for systems partially benefiting from the federation or affected by platform overhead (Medium); and for systems fully benefiting from the federation to conduct realistic experimentation with other testbeds (High).
\end{itemize}

Ideally, the experimental infrastructure should have high controllability, observability, repeatability, and fidelity. Such high-performance measures are only possible in real physical testbeds that are co-located, syntactically compatible, and managed by the same entity; an exercise that is cost-prohibitive and difficult to scale. Thus, an experimental infrastructure similar to the FCTaaS, increases the experimental scalability through the federation, at a compromise with Controllability, Observability, Repeatability, and Fidelity. From low to high, the ratings range from 1 to 5 for controllability, 1 to 4 for observability, and 1 to 3 for repeatability and fidelity, yielding a maximum aggregate score of 15. Scores are assigned based on observations from each case study and summarize the combined operational suitability of the participating testbeds for federated CPS-security experimentation.

\section{Experimental Results} \label{sect:s4}
This section evaluates the FCTaaS framework, measuring its effectiveness and performance while performing federated experimentation. During this evaluation, we integrate (1) Smart Water testbed at the University of Arizona; (2) Smart car testbed at the University of Arizona; (3) ScadaBR testbed on Amazon Web Services (AWS); (4) Virtual Cybersecurity Attacker Testbed on Scaleway (France); (5) Suricata Testbed on AWS; (6) Ganglia Testbed on AWS; into the FCTaaS. Using these six testbeds, we stress-test the FCTaaS by measuring its performance while supporting resource-intensive attacks like Denial of Service attacks on the testbeds. This experimentation allows measuring the controllability, observability, repeatability, and fidelity of the proposed FCTaaS.

\subsection{Case Study 1: Denial of Service (DoS) Attack Experiment on Smart Water System Testbed in~FCTaaS} \label{sect:s4dot1}

This case study aims to evaluate the effects of Denial-of-Service (DoS) Attacks on critical infrastructures such as Smart Water Systems using the FCTaaS. The study leverages the following components:

\begin{itemize}
    \item \textbf{Smart Water Systems Testbed at University of Arizona:} To investigate Cyber-Physical Systems (CPS) security, the NSF IUCRC Center for Cloud and Autonomic Computing at the University of Arizona has developed a Smart Water Testbed \cite{Pacheco2017IoTSystem}. This Smart Water Testbed uses an OpenPLC-based controller \cite{THEOpenplcproject.com} to control water supply by controlling a water pump (via PLC outputs), while monitoring the water flow through water flow sensors (via PLC inputs) as illustrated in Fig.~\ref{fig:SW_TB}. Fig.~\ref{fig:SW_TB} highlights that the Smart Water Testbed is a real CPS platform rather than a purely software-based simulation. This is important for the present study because disruption of communications in such a system does not only affect network connectivity, but also affects the observation and control of the physical process. Thus, the testbed enables a more realistic evaluation of how cyber attacks influence CPS operation in the federated environment. The controller receives user control inputs from a Supervisory Control and Data Acquisition (SCADA) by listening to the Modbus protocol, enabling users to control the Smart Water Testbed.
    
    \item \textbf{ScadaBR testbed on Amazon Web Services (AWS):} The SCADA system \cite{DaneelsA.1999WHATSCADA}, helps industrial systems gather real-time data, enable intelligent decision-making, and mitigate communication inefficacies, is widely adopted by the Industry to manage CPS. ScadaBR \cite{ScadaBR} is an open-source SCADA interface that interacts with industrial data sources over different protocols such as Modbus Serial, Modbus IP, and DNP3. The CAC at the University of Arizona has a ScadaBR testbed running on the AWS cloud. Using the FCTaaS, this ScadaBR testbed will allow the FCTaaS user to control the Smart Water Systems Testbed via the Modbus protocol. 

    \item \textbf{Virtual Cybersecurity Attacker Testbed on Scaleway (France):} The Virtual Cybersecurity Attacker Testbed is a Docker container  \cite{EmpoweringDocker} deployed on Scaleway (France) with 2GB memory and two x86 64bit cores. The Virtual Cybersecurity Attacker Testbed has attacker tools to perform cybersecurity attacks for cybersecurity experimentation. In this case study, the Virtual Cybersecurity Attacker Testbed will behave as the attacker, targetting the CPS operation with a Denial of Service (DoS) attack using the~FCTaaS.
\end{itemize}

In this case study, the FCTaaS combines these three testbeds to allow an FCTaaS user to perform a cybersecurity experiment. Each testbed integrates into the FCTaaS using the REST API to join the experiment VPN. After that, the federation will allow users to take advantage of the testbed resources through a secure cloud API. This case study aims to explore research questions like: “What is the necessary DoS bandwidth for an intruder to take down a critical infrastructure service?”. Moreover, experiments like this case study help analyze the cascade effects of cyberattacks on CPS.

\subsubsection{Setting up the Experiment}\label{sect:s4dot1dot1}

Before establishing the experiment, the user first browses available testbeds in the FCTaaS repository through the Web Interface to identify relevant and available testbeds. For example, Fig.~\ref{fig:SW_Exp} depicts an anticipated scenario where the ScadaBR testbed controls the Smart Water Testbed, and the Virtual Cybersecurity Attacker Testbed is to launch attacks on the Smart Water Testbed and the ScadaBr testbed~\cite{Vuletic2018REALIZATIONLINUX}. {The figure shows that the experiment involves a distributed control path and an attack path operating simultaneously across geographically isolated testbeds. This is an important capability of the FCTaaS, as it allows the researcher to evaluate the behavior of a remotely controlled CPS while it is subjected to adversarial traffic within the same federated environment. The user begins the experiment setup by selecting the testbeds and the inputs necessary. During this selection, the user also initializes experiment configuration information, like the port number for the SCADA communication between the Smart Water Testbed and the ScadaBR testbed. After the experiment setup is complete, this configuration information is shared with the individual Testbed Managers to set up the testbed for the experiment. The testbeds are ready for the experiment when the Testbed Managers join the experiment VPN; the network on which the experiments will happen.

\begin{figure}[H]
  \centering
  \includegraphics[width=.8\columnwidth]{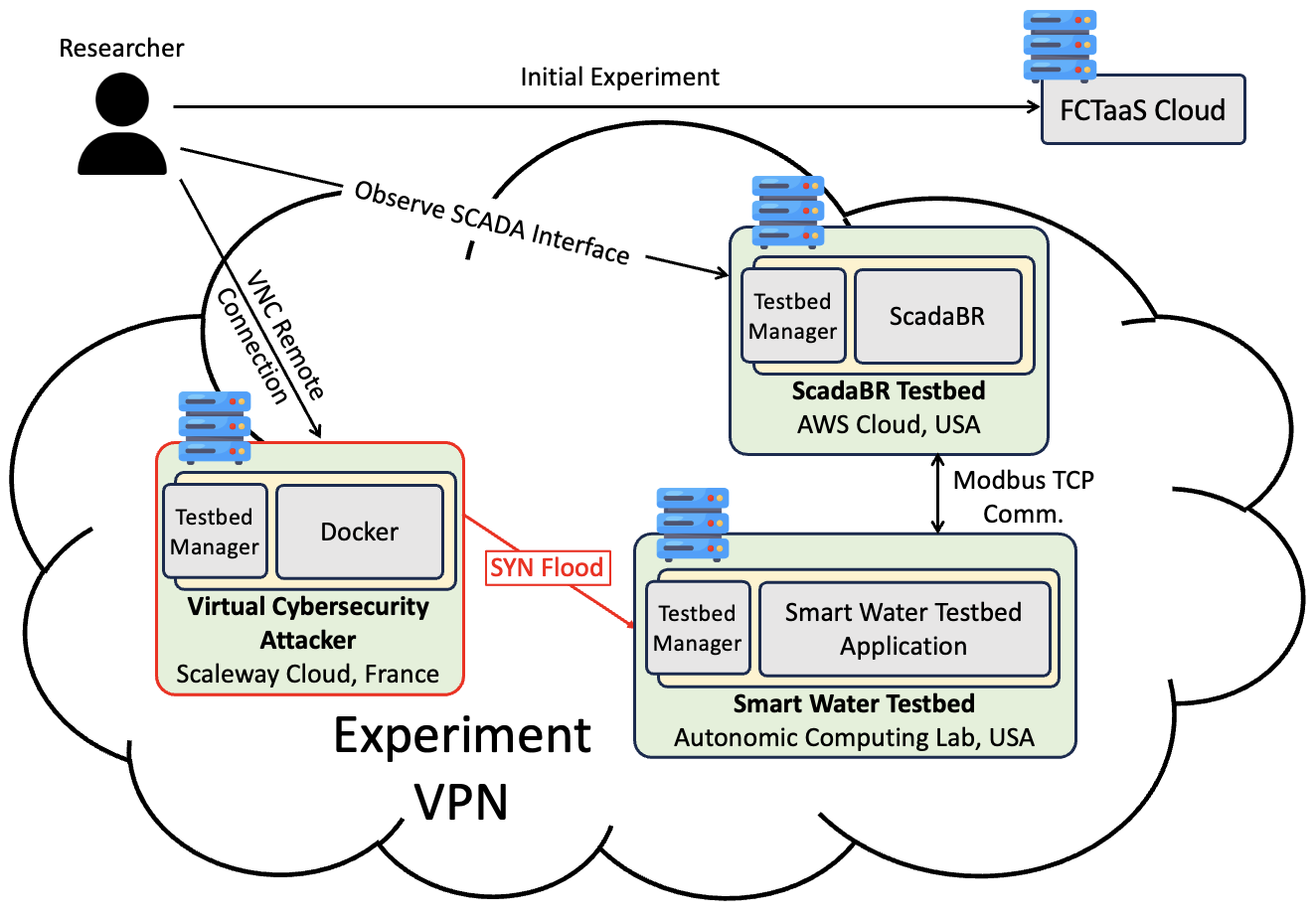}
  \caption{Case Study 1: Denial of Service Attack on Smart Water Testbed. The figure illustrates the federated composition of the ScadaBR testbed, the Virtual Cybersecurity Attacker Testbed, and the Smart Water Testbed through the experiment VPN for evaluating a distributed DoS attack scenario. }
  \label{fig:SW_Exp}
\end{figure}

The federation formation procedure adheres to the experiment creation algorithm mentioned earlier, where the FCTaaS Cloud ($C_{fct}$) first checks with the Privacy and Security service ($S_{ps}$). After receiving permission from $S_{ps}$, $C_{fct}$ forwards the request to the Testbed Manager Service ($S_{mg}$). $S_{mg}$ then sends availability requests to each testbed. If the required testbeds are available, then $S_{mg}$ returns the confirmation message to the user. 

Next, the user will start the initialization procedure through REST API calls that set the procedure by sending experiment creation requests to  $S_{mg}$. Then $S_{mg}$ will send initialization events to the selected testbeds along with their initialization parameters, including the required network configurations. The Testbed Managers associated with each selected testbed will trigger the experiment initialization event handler and join the experiment. Once this step is completed, $S_{mg}$ will pass the necessary credentials and internal or external IP addresses that are required, if applicable, to access these testbeds. The user will receive these credentials and access procedures experiment on the web interface, invoking the start of the experimentation phase.

\subsubsection{Experiment and Results}\label{sect:s4dot1dot2}

This experiment aims to assess the impact of the DoS attack against the Smart Water Testbed. The user can observe the impact of a successful attack from the ScadaBR Web interface. During regular operations (i.e., before the attack), the user can observe and control the status of the Smart Water Tesbed with the ScadaBR. The ScadaBR shows alerts complaining about the loss of connection to the Smart Water Testbed on successful DoS attack execution. These alerts will persist till the user acknowledges them. Fig.~\ref{fig:1Exp}, highlights the attack success rate versus bandwidth utilized for the attack, where ScadaBR showing loss of connection alert is considered a successful attack. The figure shows that the relationship between attack bandwidth and attack success is progressive rather than binary. At lower traffic rates, the attack is either unsuccessful or only partially successful, indicating that the communication path retains some resilience under limited stress. However, as the bandwidth increases, the attack success rate also increases, reaching full success at 5 Mbps and above. This result is important because it provides an estimate of the disruption threshold required to deny SCADA control of the Smart Water Testbed in the studied deployment.

\begin{figure}[H]
  \centering
  \includegraphics[width=0.9\columnwidth]{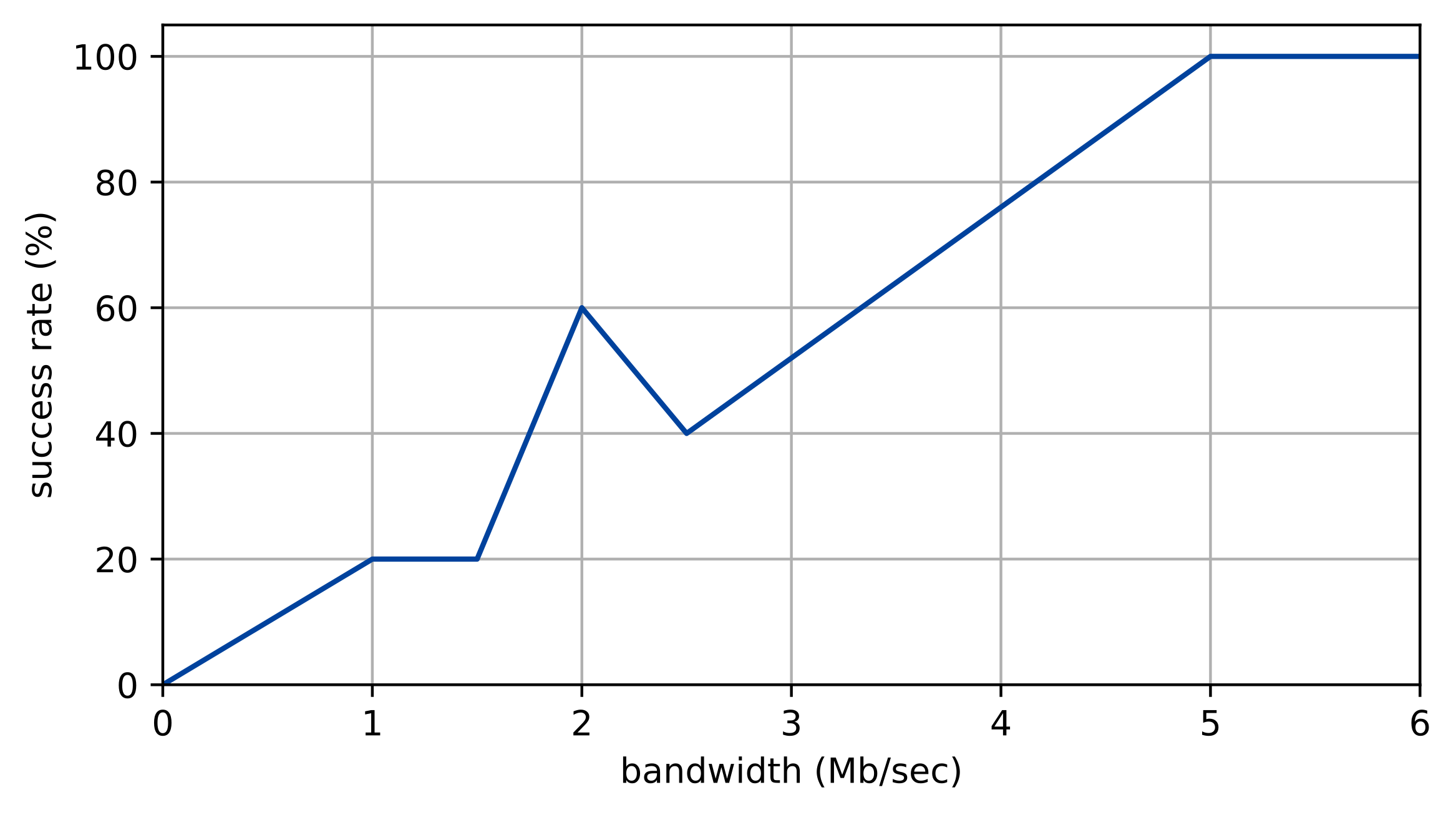}
  \caption{DoS attack targeting the Smart Water Testbed. The figure presents the attack success rate as a function of the attack bandwidth, showing the increase in attack effectiveness as the bandwidth of the DoS traffic increases.}
  \label{fig:1Exp}
\end{figure}

In this scenario, the user performs DoS attacks five times at each of the following rates: 0.5, 1, 1.5, 2, 2.5, 5, and 10 Mbps. Attacks larger than or equal to 5 Mbps successfully performed Denial of Service attacks on the Smart Water Testbed, wherein the ScadaBR was unable to control the testbed \cite{Huitsing2008AttackProtocols}. We want to highlight that such DoS attacks, while launched during experiments, can also disrupt the federation's operations. In this case study, although ScadaBR could no longer control the Smart Water Testbed through the Modbus/TCP path, the testbed remained visible to FCTaaS through heartbeat messages generated by its Testbed Manager. This behavior follows the intended FCTaaS design, in which testbed-level liveness reporting is logically separated from application-level control communication and does not depend on successful Modbus/TCP transactions. Therefore, the heartbeat confirmed that the Testbed Manager and testbed remained online, although the SCADA-controlled application was temporarily unreachable.

Lastly, we perform a latency analysis, comparing the execution latency of the same experiment when performed in the FCTaaS, versus using co-located testbeds in the same local area network (LAN). Our experimental results show a delay of less than 1 ms while using a co-located testbed in the same LAN, while an average delay of 137 ms when using the FCTaaS. The delay introduced while using the FCTaaS is due to networking delay where, two testbeds are in the US (one on AWS cloud), and the third testbed is in France. Moreover, this average latency of 137ms still allows for the successful completion of the experiment, not hampering the observed results. 

Upon completing the experiment, the user terminates the experiment with a termination request sent to the Experiment Management Service through the web interface. The Experiment Management Service triggers the Testbed Manager to perform experiment finalization, making the testbed available for other experiments on the FCTaaS.

\subsubsection{Case Study 1: Performance Evaluation Metrics:}\label{sect:s4dot1dot3}
In evaluating the performance metrics of the testbeds utilized in the Denial of Service (DoS) Attack Experiment on the Smart Water System Testbed in FCTaaS, various aspects were taken into account. Firstly, the Smart Water Testbed demonstrated moderate controllability, enabling users to manage water supply and monitor flow through sensors, albeit with limitations compared to more advanced systems. Observability was also moderate, providing users with insights into system status and alerts during both normal operations and attack scenarios, although enhancements in data granularity could be beneficial. The testbed's repeatability scored fairly, allowing experiments to be replicated to a certain extent, though challenges in reproducing specific environmental conditions were noted. However, fidelity was relatively low due to constraints in accurately simulating real-world scenarios, potentially impacting its ability to replicate complex interactions and responses. Overall, the Smart Water Testbed attained a score of 9, indicating its suitability for experimentation but acknowledging certain limitations. 

Conversely, the ScadaBR Testbed exhibited high controllability, enabling effective interaction with industrial data sources and CPS operations. Observability was also high, facilitating real-time monitoring of system status and prompt detection of anomalies. Repeatability, although moderate, allowed for experiments to be replicated to a certain extent, despite potential variations in network conditions or external factors. Moreover, fidelity was relatively high, as the testbed could simulate realistic industrial scenarios and responses, bolstering the credibility of experimental results. Consequently, the ScadaBR Testbed garnered an overall score of 12, highlighting its significant suitability and effectiveness for conducting experiments in this context.

Meanwhile, the Virtual Cybersecurity Attacker Testbed demonstrated moderate controllability, enabling users to execute cybersecurity attacks and manipulate parameters within a Docker container environment. Observability was high, facilitating effective monitoring of attack execution and analysis of impacts on target systems. Repeatability, although moderate, allowed for the replication of attacks to a certain extent, notwithstanding potential variations in attack conditions or system responses. However, fidelity was relatively low, as the testbed may not fully replicate the complexities of real-world cyberattacks and their effects on critical infrastructure. 

Consequently, the Virtual Cybersecurity Attacker Testbed achieved an overall score of 10, indicating its moderate suitability for conducting cybersecurity experiments. Collectively, with an average score of approximately 10.33, the case study underscores the effectiveness and suitability of the testbeds for conducting experiments related to cybersecurity and critical infrastructure.

\subsection{Case Study 2: Detection and Analysis of DoS Attack on the Smart Water Testbed} \label{sect:s4dot2}
This Case Study continues with the previous case study mains to address the following two research~questions:
\begin{enumerate}
  \item How can we detect such DoS attacks in critical infrastructure environments?
  \item What stress is put on intrusion detection mechanisms while undergoing a DoS attack?
\end{enumerate}

In this case study, the FCTaaS user will analyze the traffic from the Smart Water Testbed to identify the potential threats endangering the CPS systems functionality. Through this traffic analysis, the user aims to detect the attacks and take countermeasures; aiming to retain the CPS' operations. The FCTaaS user will analyze the Smart Water Testbed traffic and take countermeasures using the Suricate Testbed~\cite{Suricata}, while analyzing the network traffic and resource utilization with the Ganglia testbed as shown in Fig.~\ref{fig:2Exp}. The figure highlights that Case Study 2 extends the federation from attack execution to a broader intrusion-detection and monitoring workflow. By inserting the Suricata Testbed in the traffic path and using the Ganglia Testbed for monitoring, the FCTaaS enables the researcher to observe whether the attack is detected, whether prevention actions can be enforced, and how much stress is placed on the detection system during the experiment. Hence, the figure demonstrates that the FCTaaS supports coordinated attack, defense, and monitoring activities within a single federated scenario. This design demonstrates how FCTaaS can support end-to-end intrusion-detection experimentation, from attack generation and traffic inspection to alert validation, packet dropping, and IDS resource profiling.

Suricata is a free and open-source Intrusion Detection System (IDS) and inline Intrusion Prevention System (IPS) \cite{Suricata}. The Suricata Testbed comprises the Suricata software and TCP proxy running on a containerized environment on Amazon AWS with two 64-bit CPUs and 4 GB of RAM \cite{CloudAWS}. The TCP proxy allows the Suricata testbed to be in line and prevent cyber attacks. A user can observe the Suricata logs and warnings using a remote desktop.

Ganglia is an open-source distributed monitoring tool for capturing CPU loads and network utilization~\cite{GangliaSystem}. The Ganglia testbed in the FCTaaS experiments allows the experimenter to harvest systems resource information such as CPU, network, disk performance, and metrics, visualized in the Ganglia tool for better understanding. The Ganglia Testbed has an instance of the Ganglia tool running in a containerized environment on the AWS cloud with two 64-bit CPUs and 4 GB of RAM. Ganglia was selected due to it's low overhead when monitoring a test bed, minimizing additional resource consumption.

Through this case study, we will expand on the resource utilization analysis performed in the previous case study and evaluate the working of the FCTaaS while performing resource-intensive cyberattacks.

\subsubsection{Detecting DoS Attack} \label{sect:s4dot2dot1}

The FCTaaS user launches a SYN flood Denial of Service attack on the Smart Water Testbed using the Virtual Cybersecurity Attacker Testbed, sending over 200K packets in ninety microseconds. The attack traffic for the Smart Water Testbed is routed through the Suricata Testbed (using the TCP proxy). The Suricata Testbed detects the SYN flood attack and raises an alert. This confirms that the federated architecture can place an IDS/IPS in the traffic path, inspect attack traffic in real time, and expose detection evidence to the researcher through Suricata logs and warnings. The FCTaaS user can also set the Suricata rule to drop packets on alert, turning detection into active intrusion prevention and allowing the Smart Water Testbed to retain its operation during the attack. This confirms that the IDS effectively identified the attack, generated an actionable alert, and enforced packet drop rules to preserve system availability.

\begin{figure}[H]
  \centering
  \includegraphics[width=.8\columnwidth]{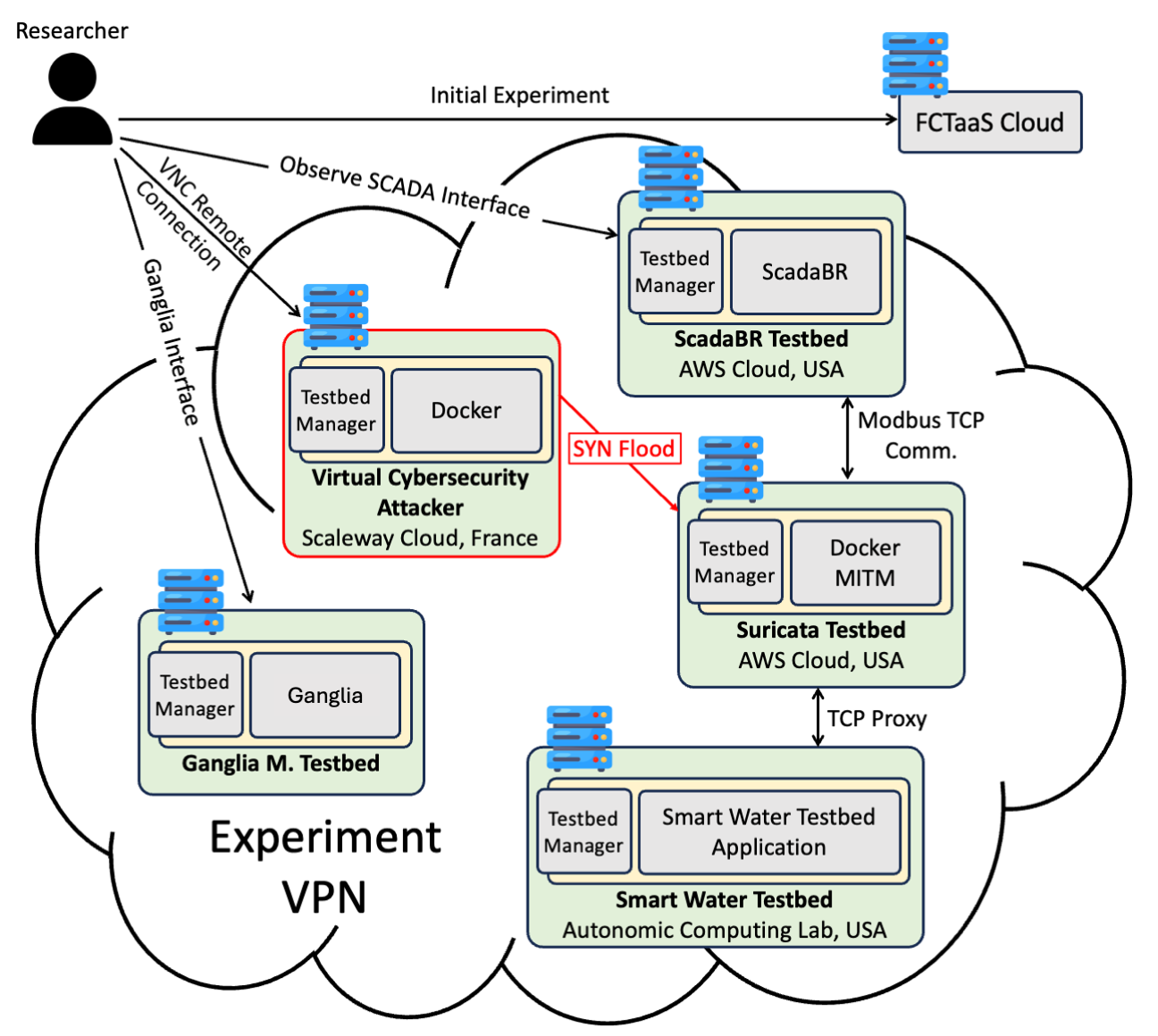}
  \caption{Case Study 2: Detecting and Analysis of DoS attack on the Smart Water Testbed. The figure shows the integration of the Suricata Testbed and the Ganglia Testbed into the federated workflow, enabling attack detection, prevention, and resource-utilization monitoring during the DoS experiment.}
  \label{fig:2Exp}
\end{figure}

\subsubsection{Intrusion Detection Stress Analysis Under DoS Attack} \label{sect:s4dot2dot2}

This section evaluates the impact of resource-intensive Denial of Service (DoS) attacks on the FCTaaS with particular emphasis on the operational stress placed on the intrusion detection layer. Although the experiment structures are the same as in Fig.~\ref{fig:2Exp}, we generate attack traffic up to 16Mbps. For the attack period of 15 min, the Suricata systems resource utilization is observed using the Ganglia testbed. The Suricata Testbed analyzes each incoming packet consuming significantly higher resources than during normal operations in its IDS role. Thus, the experiment evaluates not only whether the IDS can detect the SYN flood, but also whether the federated framework can sustain IDS monitoring while the detection engine experiences elevated CPU, memory, disk, and network load. Fig.~\ref{fig:3Exp} shows the increased CPU utilization, disk space utilization, server load, free memory, and the network utilization of the Suricata Testbed. In Fig.~\ref{fig:3Exp}, we observe significantly higher resource utilization during the attack (13:35--13:50).


\begin{figure}[H]
  \centering
  \includegraphics[width=0.9\columnwidth]{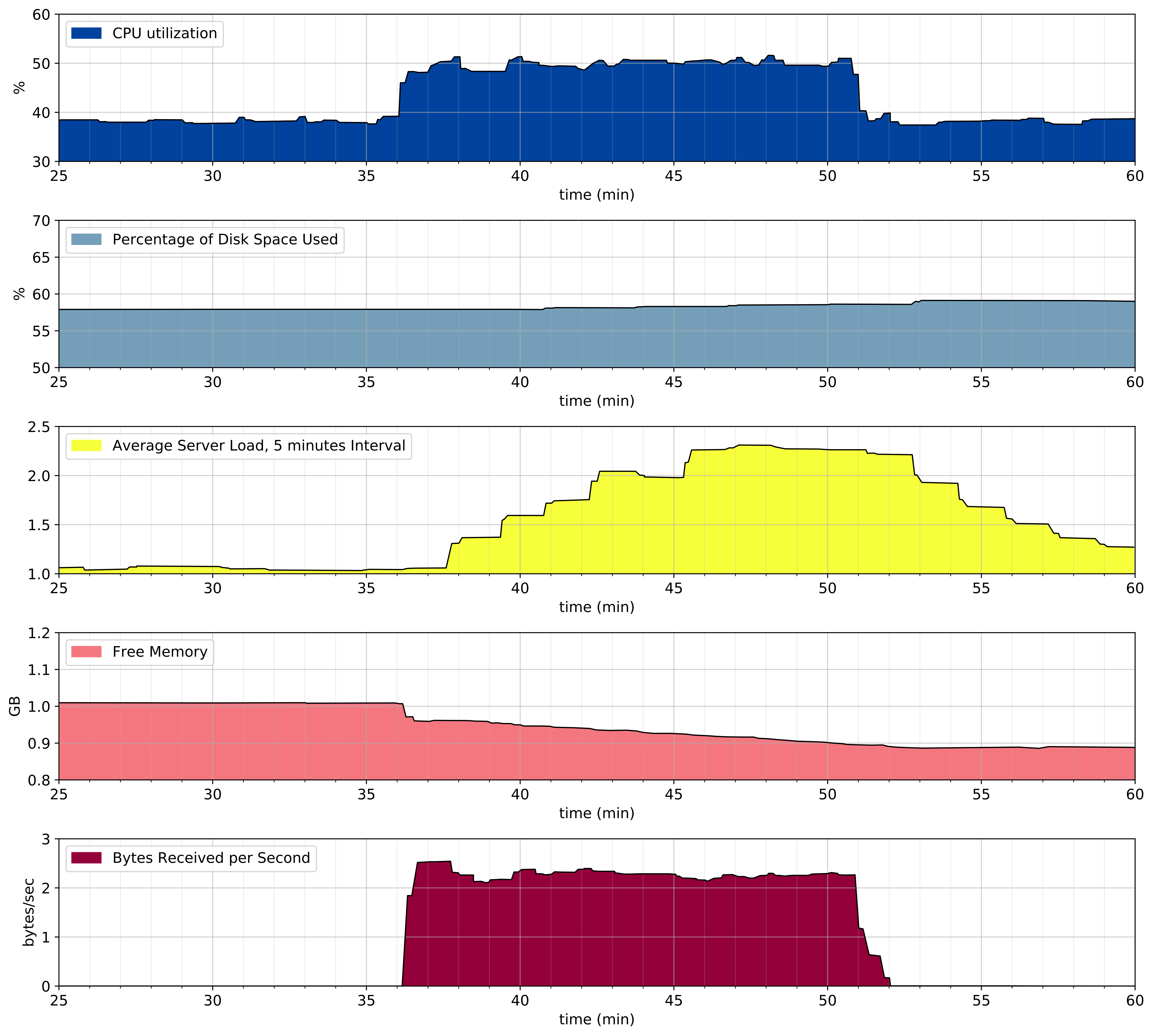}
  \caption{
  Resource utilization of the Suricata Testbed during the federated DoS experiment. From top to bottom, the plots show CPU utilization, disk utilization, five-minute average system load, available memory, and received network throughput. The shaded attack interval highlights the increased processing and network demand associated with inline IDS/IPS operation under the DoS workload.}
  \label{fig:3Exp}
\end{figure}

The data shows an average CPU utilization of 38\% during regular operation, whereas the average increases to 49\% when under attack.  Similarly, the average load, representing runnable jobs in the scheduler's run queue per unit time, increased from 1.0 to 2.3. In contrast, the attack caused a decrease in the free memory from 1 GB to 0.9 GB. Moreover, the network resource utilization jumps up from less than 2.2 Mbps to 16 Mbps during the attack, causing disk space utilization to increase from 58\% to 59\% (due to increased logs). When accessed over the remote desktop, the Suricata testbed had a longer response time while displaying during the attack. However, the Smart Water testbed, the ScadaBR testbed, and the Ganglia testbed did not show any degradation in their connections to the FCTaaS during the attack.
In conclusion, we observed a significant increase in resource utilization during this case study while performing resource-intensive attacks. However, the FCTaaS framework sustained its operation, allowing the FCTaaS user to complete their case study scenario.

\subsubsection{Case Study 2: Performance Evaluation Metrics} \label{sect:s4dot2dot3}
In the context of Case Study 2, the performance evaluation metrics of the involved testbeds are assessed to gauge their suitability and effectiveness in detecting and analyzing Denial of Service (DoS) attacks on the Smart Water Testbed within the FCTaaS framework. The Smart Water Testbed exhibits moderate controllability, enabling users to manage water supply and monitor flow through sensors, although with certain limitations compared to advanced systems. Observability is also moderate, providing users with insights into system status and alerts during normal operations and attack scenarios. Repeatability scores fair, allowing for the replication of experiments to a certain extent, albeit with challenges in reproducing specific environmental conditions. However, fidelity is relatively low due to constraints in accurately simulating real-world scenarios, impacting its ability to replicate complex interactions and responses. Overall, the Smart Water Testbed achieves a score of 9, signifying its suitability for experimentation but acknowledging certain limitations.

Meanwhile, the Suricata Testbed demonstrates high controllability and observability, allowing effective detection and prevention of cyberattacks through its Intrusion Detection System (IDS) and Intrusion Prevention System (IPS) functionalities. Repeatability is moderate, enabling experiments to be replicated, while fidelity is relatively high as the testbed can simulate realistic attack scenarios. Consequently, the Suricata Testbed garners an overall score of 12, indicating its significant suitability and effectiveness for detecting and analyzing DoS attacks.

Similarly, the ScadaBR Testbed also showcases high controllability and observability, providing users with real-time data and effective control over CPS operations. Repeatability is moderate, allowing for the replication of experiments, while fidelity is relatively high as the testbed can simulate realistic industrial scenarios. Thus, the ScadaBR Testbed attains an overall score of 12, emphasizing its high suitability and effectiveness for experimentation.

The Virtual Cybersecurity Attacker Testbed offers moderate controllability and high observability, enabling users to execute cyberattacks and monitor their impact effectively. Repeatability is moderate, allowing for the replication of attacks, while fidelity is relatively low due to limitations in simulating complex real-world cyberattacks. The testbed achieves an overall score of 10, indicating its moderate suitability for conducting cyberattack experiments.

Furthermore, the Ganglia Monitoring Testbed demonstrates high controllability and observability, providing comprehensive monitoring of system resources. Repeatability is moderate, allowing for the replication of resource utilization experiments, while fidelity is high as the testbed accurately captures resource metrics. Consequently, the Ganglia Testbed achieves an overall score of 12, underscoring its high suitability and effectiveness for monitoring system resource utilization.

Overall, Case Study 2 achieves an average score of 11, reflecting the collective effectiveness and suitability of the testbeds in detecting and analyzing DoS attacks on critical infrastructure within the FCTaaS framework. Also, the results show that FCTaaS can support intrusion detection experimentation as a first-class use case where researchers can deploy an IDS/IPS in line, validate alerts, trigger packet-dropping rules, and quantify the resource burden imposed on detection mechanisms during high-volume attack.

\subsection{Case Study 3: Man in the Middle (MITM) Attack Evaluation on Smart Car Testbed} \label{sect:s4dot3}

The FCTaaS user investigates man-in-the-middle attacks on Autonomous Vehicles through this case study. As illustrated in Fig.~\ref{fig:4Exp}, this case study uses the Virtual Cybersecurity Attacker Testbed, Smart Car testbed located at the University of Detroit Mercy, and Remote Server Testbed located on Scaleway in France. The figure shows that Case Study 3 evaluates a different class of cybersecurity threat from the DoS-based case studies. Instead of targeting service availability, the MITM workflow targets the integrity of the communications between the Smart Car Testbed and the Remote Server Testbed. This is important because it demonstrates that the FCTaaS can support not only traffic-flooding attacks, but also experiments involving communication interception, packet manipulation, and data injection across geographically distributed CPS components.

The Smart Car testbed, located at the University of Arizona, consists of a model car controlled by a controller. A user controls this car using wireless telemetry modules, WiFi, and Bluetooth, and the attacker aims to exploit these communication links to perform the MITM attack \cite{WenPlug-N-Pwned:IoT}.   

The Remote Server testbed is a simulated vehicle control center deployed on Scaleway in France, that receives the telemetry data from the vehicles. The Remote Server analyzes the live car data to observe the driving conditions and behavior.

\begin{figure}[H]
  \centering
  \includegraphics[width=.8\columnwidth]{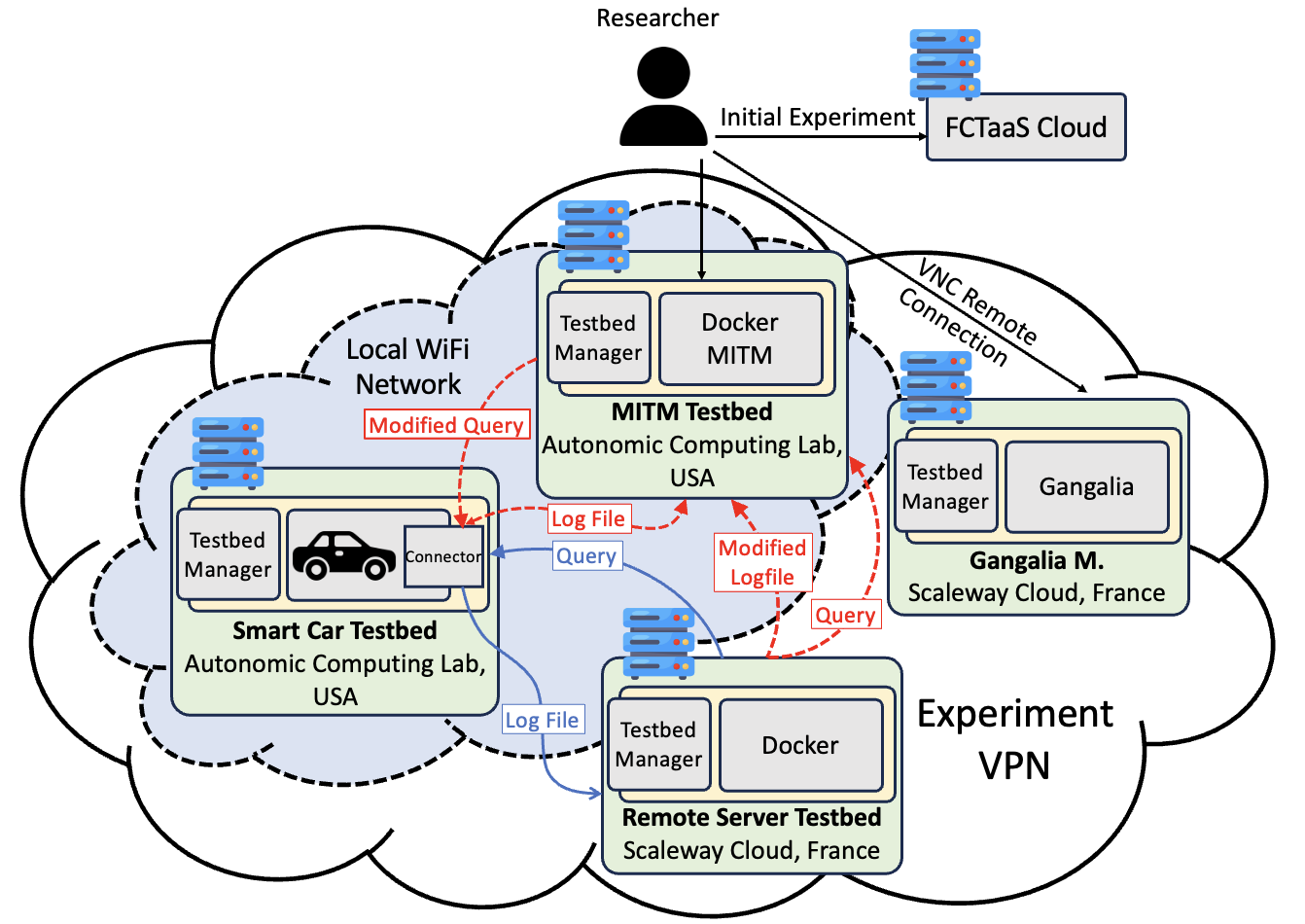}
  \caption{Case Study 3: Man in the Middle Attack (MITM) evaluation on Smart Car Testbed. The figure illustrates the federated setup used to intercept and manipulate communications between the Smart Car Testbed and the Remote Server Testbed while monitoring resource utilization through the Ganglia Testbed.}
  \label{fig:4Exp}
\end{figure}

\subsubsection{Setting up the Experiment} \label{sect:s4dot3dot1}

The FCTaaS user aims to execute a MITM attack using the Virtual Cybersecurity Attacker Testbed, by exploiting the communications between the Remote Server Testbed and the Smart Car Testbed. During the MITM attack, the user will intercept the Car and the Remote Server's communications and inject its messages into the communications. Each testbed has its Testbed Managers that facilitate the integration of the testbeds into the federation. 

\subsubsection{Experiment and Results} \label{sect:s4dot3dot2}

The MITM attack aims to intercept communication packets between the Smart Car Testbed and the Remote Server Testbed. To perform a successful MITM attack, the FCTaaS user attacks the WiFi communications used by the Smart Car Testbed, by hosting an evil WiFi access point (AP) \cite{Sheldon2012TheNetworks, Aircrack-ngAireplay-ng,satam2020wids}, with the same name as the real communication AP. On achieving an MITM, the attacker uses Bettercap \cite{BETTERCAP}, a network sniffing and traffic capturing tool to sniff embedded tokens in the communications. The attacker starts injecting fake data into the communication stream with these embedded tokens, creating potentially life-threatening situations. The FCTaaS user uses the Ganglia Testbed to observe the network, CPU, and memory utilization throughout the experiment. Similar to case studies 1, and 2, the attacker's use of Bettercap increases the resource utilization of the Virtual Cybersecurity Attacker Testbed drastically, as shown in Fig.~\ref{fig:5Exp} at 18:09:24. The figure shows that the attacker-side workload during the MITM experiment is primarily reflected in CPU utilization. The fluctuations in BetterCap CPU utilization indicate the changing processing demand associated with packet sniffing, traffic inspection, and message injection during the attack. In contrast, the memory utilization remains comparatively stable throughout the observation interval. This result suggests that, in the studied scenario, the MITM attack is more sensitive to real-time traffic-processing demand than to memory consumption, thereby providing additional insight into the computational behavior of the attacker testbed in the federated environment.

\subsubsection{Case Study 3: Performance Evaluation Metrics} \label{sect:s4dot3dot3}

In Case Study 3, the performance evaluation metrics of the testbeds involved in evaluating Man-in-the-Middle (MITM) attacks on Smart Car Testbed are assessed to understand their effectiveness in detecting and mitigating such attacks within the FCTaaS framework.

The Smart Car Testbed, located at the University of Detroit Mercy, demonstrates moderate controllability, enabling users to control the model car via wireless telemetry modules, WiFi, and Bluetooth. Observability is also moderate, allowing users to monitor the behavior of the car during normal operations and attack scenarios. However, repeatability is fair, indicating challenges in precisely reproducing certain experimental conditions. Additionally, fidelity is relatively low due to limitations in accurately simulating real-world scenarios. Overall, the Smart Car Testbed achieves a score of 9, indicating its suitability for experimenting with MITM attacks but acknowledging certain limitations.

Meanwhile, the MITM Testbed exhibits high controllability and observability, allowing users to intercept and manipulate communications between the Smart Car Testbed and the Remote Server Testbed effectively. Repeatability is moderate, enabling the replication of MITM attack scenarios, while fidelity is relatively high as the testbed can accurately simulate MITM attack scenarios. Consequently, the MITM Testbed garners an overall score of 12, emphasizing its significant suitability and effectiveness for evaluating MITM attacks.

Similarly, the Ganglia Monitoring Testbed demonstrates moderate controllability and high observability, providing comprehensive monitoring of system resources throughout the experiment. Repeatability is fair, allowing for the replication of resource utilization experiments, while fidelity is relatively low due to limitations in accurately simulating real-world resource utilization scenarios. The Ganglia Testbed achieves an overall score of 10, indicating its moderate suitability for monitoring resource utilization during the~experiment.

Furthermore, the Remote Server Testbed showcases moderate controllability and high observability, allowing users to analyze telemetry data from the Smart Car Testbed effectively. Repeatability is fair, enabling the replication of experiment scenarios, while fidelity is relatively low due to limitations in accurately simulating real-world driving conditions. The Remote Server Testbed achieves an overall score of 10, indicating its moderate suitability for analyzing Smart Car Testbed data.

Overall, Case Study 3 achieves an average score of 10.25, reflecting the collective effectiveness and suitability of the testbeds in evaluating MITM attacks on Smart Car Testbed within the FCTaaS framework.

\begin{figure}[H]
  \centering
  \includegraphics[width=0.8\columnwidth]{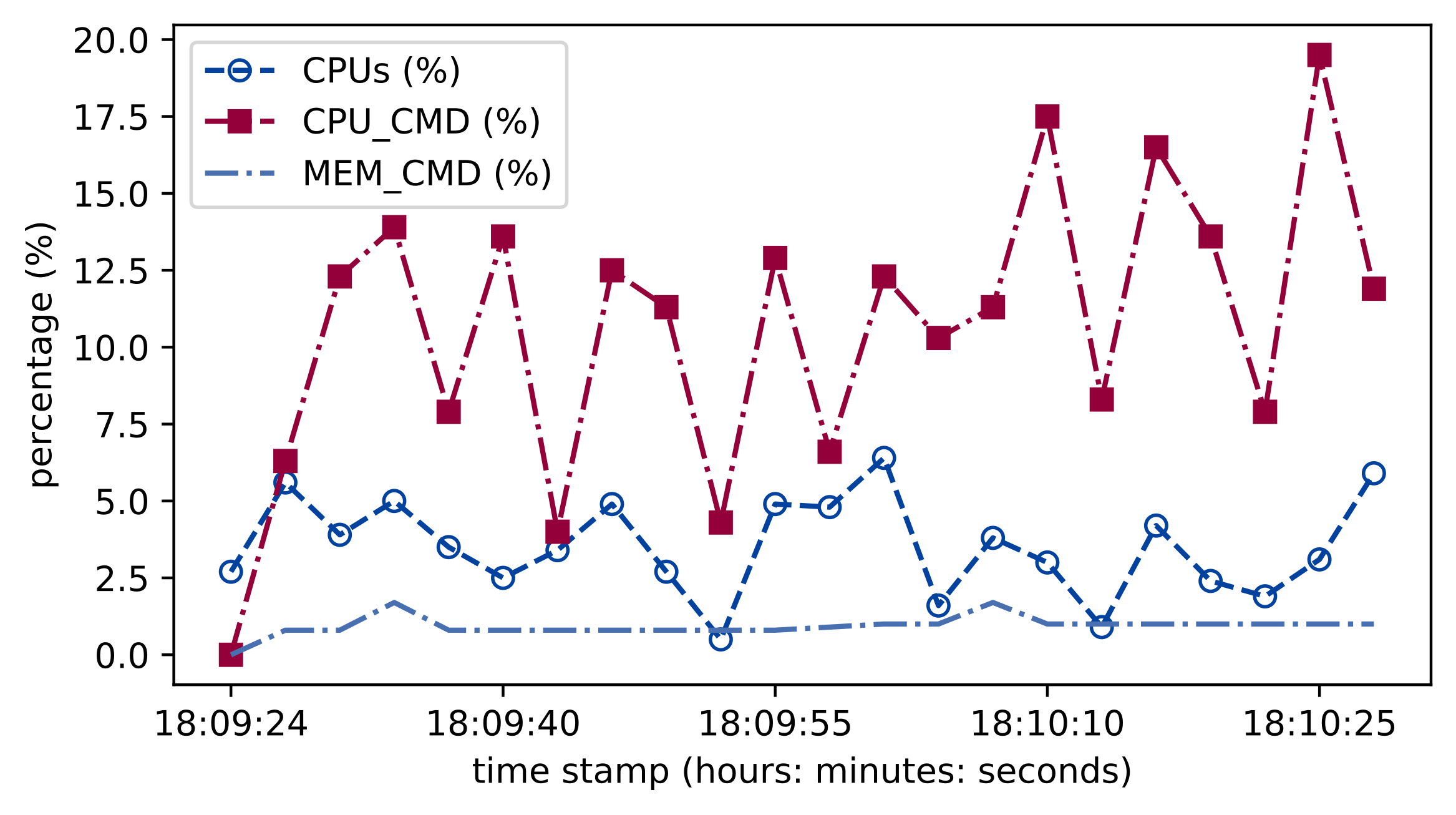}
  \caption{Overall CPU utilization, BetterCap CPU utilization, and BetterCap memory utilization. The figure shows the resource utilization behavior of the attacker-side environment during the MITM experiment, highlighting the processing overhead associated with active interception and traffic manipulation.}
  \label{fig:5Exp}
\end{figure}

\subsection{Federation Overhead Analysis} \label{sec:overheadanalysis} \label{sect:s4dot4}

The FCTaaS framework allows for better and more efficient utilization of expensive hardware, allows sharing of resources, and provides an avenue for better experimentation. FCTaaS also reduces experiment setup time while simulating complex experimental scenarios and assigning small overhead costs to the participating testbeds. This section aims to quantify the overhead associated with integrating a testbed into the FCTaaS.  To better understand the recurring overhead, we disregard the one-time cost necessary to set up the testbed, such as hardware cost, time to assemble, programming cost, and time required to integrate into the framework. For this analysis, we focus on Case Study 1, where the Smart Water Testbed has an OpenPLC \cite{THEOpenplcproject.com} controller built on a Raspberry Pi 4 Model B with a 1.5 GHz 64-bit quad-core ARM Cortex-A72 processor and 4 GB of RAM.

Various data were gathered from the testbed with and without the federation to measure the framework overhead cost. The experiment ran for ten minutes, collecting overhead observations every 15 s while testbeds executed their respective individual experimentation. In total, 34 data samples in the local network and 34 data samples in the federation were collected, highlighting device resources such as CPU and Memory Utilization. The goal is to establish a baseline for an experiment running on an isolated testbed, thereby comparing results with data collected after integrating the FCTaaS framework. Experimenting without the federated environment has an average number of 178.7 tasks running with an overhead of 5.45\% on CPU and 9.09\% on RAM, as presented in Table~\ref{tab:overhead_analysis}. The table shows that the recurring overhead introduced by the federation is limited relative to the baseline resource usage of the testbed. After integration into the FCTaaS framework, the average number of tasks increases only slightly from 178.70 to 181.03, while the CPU utilization increases by 0.29\% and memory utilization increases by 0.99\%. These results indicate that the additional services required for remote access, coordination, and secure experiment execution impose only a minor resource cost on the host platform. Therefore, the table supports the claim that the FCTaaS enables federated experimentation without significantly degrading the performance of the participating~testbed.

\begin{table}[H]
\caption{Federation Overhead Analysis. The table compares the Smart Water Testbed resource utilization with and without federation, showing that integration into the FCTaaS introduces only a small recurring increase in tasks, CPU utilization, and memory usage.}
\label{tab:overhead_analysis}
\centering
\begin{tabular}{ccccc}
\hline
                  & \textbf{Avg. Total Tasks} & \textbf{Avg. CPU} (\%) & \textbf{Memory Used} (Mibs) & \textbf{Avg. Memory} (\%)  \\ \hline
Non Federated     & 178.70                    & 5.45                   & 348.81                      & 9.09  \\
Federated         & 181.03                    & 5.74                   & 386.76                      & 10.08  \\
\hline
Overhead          & 2.32                      & 0.29                   & 37.95                       & 0.99  \\
\hline
\end{tabular}
\end{table}

On the other hand, after integrating into the FCTaaS framework, the average number of tasks increased to 181.03. The slight increase in the number of tasks led to more consumption of the device resources. As a result, the federated model led to an average overhead of 2.32 in the number of tasks running, increasing CPU overhead by 0.29\% and 0.99\% on RAM usage. This behavior is due to the functions running on the device to enable remote access to the testbed through a secure framework. We observe an overall overhead of less than 1\% on host resources throughout the analysis. The 34-sample dataset collected over a single 10-min run provides a consistent point estimate of federation overhead under this configuration; formal statistical characterization across multiple independent runs and varying loads is acknowledged as a limitation of the current analysis.

Based on the location of the testbed and the user, a notable variation in the network latency is observed. As a result, we measure the round trip time from the different locations used in this experiment: local (Arizona), Oregon (US), and France. We collected packet latency data from these different locations and observed that the testbeds located in France had an average latency of 147.07 ms, 47.80 ms for the testbeds in Oregon, and a latency of 5.63 ms for the local testbeds. Considering the significant advantages that the model enables, we conclude that the FCTaaS framework does not add significant overhead on the testbed resources and does not hinder performance while executing an experiment. Thus, the federation provides an excellent foundation for utilizing cyber testbeds and creating a federated environment.

\subsection{FCTaaS Performance Analysis} \label{Sec:PerfAnalysis} \label{sect:s4dot5}

This subsection evaluates the controllability, observability, repeatability, and fidelity of the FCTaaS with respect to the three case studies presented. The summarized scores for the three study cases are presented in Table \ref{tab:summarized}. The table shows that the FCTaaS achieves consistent overall performance across all three case studies, with average scores of 10.33, 11, and 10.25, respectively. A clear pattern observed in the table is that observability remains comparatively strong across the study cases, while fidelity remains the most constrained metric. This behavior is consistent with the experimental observations, where the framework preserves useful visibility into the testbeds and attack behavior, but the geographically distributed nature of the federation introduces latency that limits the realism of time-sensitive interactions. Thus, providing a concise summary of the main tradeoff in the FCTaaS: improved experimental reach and integration capability at the cost of some reduction in controllability and fidelity.

\begin{table}[H]
\caption{Federation platforms performance calculation metrics for each use case presented. The table summarizes the controllability, observability, repeatability, and fidelity scores for the testbeds used in each case study, and shows the overall performance trend of the FCTaaS across the three experimental scenarios.}
\label{tab:summarized}
\resizebox{\textwidth}{!}{%
\begin{tabular}{llccccccc}
\hline
\multicolumn{3}{l}{} &
\textbf{Controllability} &
\textbf{Observability} &
\textbf{Repeatability} &
\textbf{Fidelity} &
\textbf{Total Score} &
\textbf{Average Score of Study Case} \\ \hline
 & 
  \multicolumn{2}{l}{\begin{tabular}[c]{@{}l@{}}Smart Water Testbed\end{tabular}} &
  3 & 3 & 2 & 1 & 9 &  \\ \cline{2-8}
 & 
  \multicolumn{2}{l}{\begin{tabular}[c]{@{}l@{}}ScadaBR Testbed\end{tabular}} &
  4 & 4 & 2 & 2 & 12 &  \\ \cline{2-8}
\multirow{-3}{*}{\textbf{\begin{tabular}[c]{@{}l@{}}Case \\ Study 1\end{tabular}}} &
  \multicolumn{2}{l}{\begin{tabular}[c]{@{}l@{}}Virtual Cybersecurity\\ Attacker\end{tabular}} &
  3 & 4 & 2 & 1 & 10 & \multirow{-3}{*}{10.33} \\ \hline
 & 
  \multicolumn{2}{l}{\begin{tabular}[c]{@{}l@{}}Smart Water Testbed\end{tabular}} &
  3 & 3 & 2 & 1 & 9 &  \\ \cline{2-8}
 & 
  \multicolumn{2}{l}{\begin{tabular}[c]{@{}l@{}}Suricata Testbed\end{tabular}} &
  4 & 4 & 2 & 2 & 12 &  \\ \cline{2-8}
 & 
  \multicolumn{2}{l}{\begin{tabular}[c]{@{}l@{}}ScadaBR Testbed\end{tabular}} &
  4 & 4 & 2 & 2 & 12 &  \\ \cline{2-8}
 & 
  \multicolumn{2}{l}{\begin{tabular}[c]{@{}l@{}}Virtual Cybersecurity\\ Attacker\end{tabular}} &
  3 & 4 & 2 & 1 & 10 &  \\ \cline{2-8}
\multirow{-6}{*}{\textbf{\begin{tabular}[c]{@{}l@{}}Case\\ Study 2\end{tabular}}} &
  \multicolumn{2}{l}{\begin{tabular}[c]{@{}l@{}}Ganglia M. Testbed\end{tabular}} &
  4 & 4 & 2 & 2 & 12 & \multirow{-6}{*}{11} \\ \hline
 & 
  \multicolumn{2}{l}{\begin{tabular}[c]{@{}l@{}}Smart Car Testbed\end{tabular}} &
  3 & 3 & 2 & 1 & 9 &  \\ \cline{2-8}
 & 
  \multicolumn{2}{l}{MITM Testbed} &
  4 & 4 & 2 & 2 & 12 &  \\ \cline{2-8}
 & 
  \multicolumn{2}{l}{\begin{tabular}[c]{@{}l@{}}Ganglia M. Testbed\end{tabular}} &
  3 & 4 & 2 & 1 & 10 &  \\ \cline{2-8}
\multirow{-4}{*}{\textbf{\begin{tabular}[c]{@{}l@{}}Case\\ Study 3\end{tabular}}} &
  \multicolumn{2}{l}{\begin{tabular}[c]{@{}l@{}}Remote Server Testbed\end{tabular}} &
  3 & 4 & 2 & 1 & 10 & \multirow{-4}{*}{10.25} \\ \hline
\end{tabular}%
}
\end{table}

\subsubsection{Controllability} \label{sect:s4dot5dot1}

The experimental results for the three case studies demonstrate that the FCTaaS has a low/moderate controllability depending on the experimental scenario. The primary reason for the low is due to communication delays introduced by the location of the testbeds. For instance, in Case Study 1, the communication delay increases by 137 \% when the testbed is not co-located.

\subsubsection{Observability} \label{sect:s4dot5dot2}
The case studies show that the FCTaaS has a high observability. This high observability is attributed to the low CPU and Memory overhead of the FCTaaS, allowing accurate measurements (when not affected by the controllability) as demonstrated by the overhead analysis in Section \ref{sec:overheadanalysis}.

\subsubsection{Repeatability} \label{sect:s4dot5dot3}
The FCTaaS exhibits moderate repeatability, enabling experiments to be replicated to a certain degree. However, challenges in reproducing specific environmental conditions were observed, primarily influenced by fidelity. The simulation of real-world results poses challenges, potentially affecting its capability to replicate intricate interactions and responses.

\subsubsection{Fidelity} \label{sect:s4dot5dot4}
The findings from the three case studies underscore that the FCTaaS exhibits diminished fidelity. Similar to the controllability aspect, the reduced fidelity can be traced back to the delays introduced by communication factors. This fidelity deficiency becomes particularly pronounced when time sensitivity is crucial for experimental outcomes.

The limitations in fidelity are notably highlighted when considering time-critical experimental situations. In such instances, the impact of communication-induced delays becomes more pronounced, further compromising the FCTaaS's ability to faithfully reproduce real-world conditions and responses.

\section{Discussion} \label{sect:s5}

\subsection{Scalability and Deployment Considerations} \label{sect:s2dot1}
The current evaluation demonstrates the feasibility of FCTaaS for integrating and orchestrating heterogeneous CPS-security testbeds across geographically distributed deployments. Specifically, the study incorporates six individual testbeds with distinct roles, spanning physical CPS assets, a cloud-hosted SCADA platform, a virtual attacker environment, an IDS/IPS component, and a monitoring service. These testbeds are composed into federated DoS, intrusion-detection, and MITM workflows, demonstrating the ability of FCTaaS to coordinate complementary cyber, control, and monitoring functions within a common~experiment.

This feasibility is supported by an architecture designed for incremental testbed integration. Through the Testbed Manager and REST-based interfaces, a new testbed can be onboarded by supplying its testbed description, access-policy configuration, and initialization logic, without requiring modifications to testbeds that have already been integrated. Nevertheless, the present evaluation should not be interpreted as a large-scale scalability benchmark. In particular, the performance of concurrent multi-user and multi-testbed federations, broker-level behavior under higher coordination loads, and VPN-management efficiency at larger deployment scales remains to be systematically evaluated.

In addition to scale, the operational setting of the current experiments should also be considered when interpreting the results. The case studies were conducted in controlled CPS-security environments to enable reproducible and safe evaluation of federated attack, detection, prevention, and monitoring workflows. Although these scenarios incorporate representative physical and cyber components, including PLC-based control, SCADA communication, attack generation, IDS/IPS enforcement, and resource monitoring, they do not constitute validation in a large-scale production industrial environment. Accordingly, the results demonstrate the feasibility of FCTaaS for federating heterogeneous CPS-security testbeds, rather than certifying deployment readiness for operational industrial facilities. Evaluating FCTaaS under production-scale conditions, including site-specific safety constraints, organizational governance, vendor heterogeneity, and long-term operational requirements, remains an important direction for future work.

Finally, resource consumption in a federated deployment is inherently testbed-specific. Physical CPS assets, cloud-hosted control services, attacker environments, IDS/IPS components, and monitoring services differ in hardware capacity, software stacks, experimental roles, and traffic-processing workloads. Therefore, the resource measurements reported in this work should be interpreted as deployment- and workload-specific observations rather than directly comparable baseline profiles across all participating testbeds. A systematic multi-testbed analysis would require synchronized measurements under native idle, federation-connected, and active-experiment conditions, together with normalization for hardware capacity, traffic volume, and security-tool configuration. Such role-aware profiling remains important future work for characterizing aggregate resource behavior and potential resource contention in larger concurrent federations.

\subsection{Security Model and Considerations} \label{sect:s5dot2}
Federating independently managed testbeds expands the trust boundary of an experiment by connecting distributed infrastructure, management interfaces, control services, and experimental data paths. The primary assets considered in FCTaaS include testbed availability, experimental configurations, credentials, control interfaces, telemetry collected, and traffic exchanged among participating testbeds. The security model focuses on limiting unauthorized participation in a federated experiment and protecting experiment traffic during cross-site communication.

FCTaaS considers threats in which an unauthorized user attempts to access a testbed, an unapproved testbed attempts to join an experiment, or a network adversary attempts to observe or modify traffic exchanged across geographically distributed sites. During experiment creation, the Experiment Management Service coordinates authorization checks with the Privacy and Security Service and the relevant Testbed Managers. Testbed owners define access constraints through Access Policy Files, including authorized user groups and testbed availability windows. A testbed is included in an experiment only after the corresponding authorization and availability checks are completed. Following approval, the selected testbeds receive experiment-specific configuration and connectivity parameters. The participating testbeds then join an experiment-specific VPN overlay, which provides encrypted communication for the configured experimental data path. Remote experiment access is therefore scoped by the authorized user role, the availability policy of each participating testbed, and the credentials issued during experiment initialization. Testbed Managers remain responsible for applying the experiment configuration and reporting testbed status to the federation-management services. These mechanisms provide federation-level authorization, access scoping, and protected experiment connectivity. However, they do not constitute a complete security-hardening solution for all deployment environments. The current framework does not claim to prevent threats arising from compromised participating endpoints, stolen credentials, insecure VPN configurations, malicious authorized users, management-plane vulnerabilities, data exfiltration, or insider misuse. In addition, the IDS/IPS and monitoring components in selected case studies provide experiment-specific security functionality and are not assumed to universally protect all federation components.

Future work will investigate stronger security controls for multi-site federations, including credential lifecycle management and revocation, audit logging, endpoint attestation, finer-grained data access control, management plane hardening, and systematic security evaluation against threats targeting VPN endpoints and participating testbeds.

\subsection{Threat Coverage} \label{sect:s5dot3}
The current case studies focus on DoS/IDS-IPS and MITM scenarios because they exercise two distinct and practically relevant CPS-security workflows. The DoS/IDS-IPS case study evaluates availability-oriented threats by coordinating attack generation, inline traffic inspection and prevention, alert generation, and resource monitoring across federated testbeds. The MITM case study evaluates communication-integrity risks by demonstrating distributed traffic interception and data injection between a CPS testbed and a remote service. Together, these scenarios validate that FCTaaS can coordinate heterogeneous attacker, CPS, control, IDS/IPS, and monitoring components within a common experimental workflow.

Nevertheless, the evaluation does not provide exhaustive coverage of the cybersecurity threat landscape. In particular, the current scenarios do not evaluate threats that require persistent endpoint compromise, long-duration adversarial behavior, insider privileges, or previously unknown vulnerabilities, such as ransomware, advanced persistent threats, insider attacks, and zero-day exploits. These threat classes entail additional experimental requirements beyond those in the current case studies, including endpoint and application-layer telemetry, identity and privilege modeling, long-horizon experiment orchestration, malware-safe containment, and recovery-oriented operational measurements.

Accordingly, the present results should be interpreted as demonstrating the feasibility of FCTaaS for federating representative CPS-security experiments focused on availability and integrity, rather than establishing comprehensive threat coverage. Extending the framework to support broader threat portfolios and multi-stage adversarial campaigns remains an important direction for future work.

\subsection{Reproducibility and Artifact Availability}
FCTaaS is a deployment-oriented federation framework whose complete evaluation depends on geographically distributed physical CPS testbeds, cloud-hosted services, testbed-owner access policies, experiment-specific VPN connectivity, and deployment-specific security configurations. Therefore, a fully executable public artifact cannot be produced by releasing source code alone, and public disclosure of operational credentials, access policies, and network configurations may expose participating testbeds to unnecessary security risks. To support methodological reproducibility, this paper describes the FCTaaS architecture, the experimental lifecycle, the testbed integration mechanism, the case-study workflows, the monitoring components, and the evaluation tools. Researchers interested in reproducing or extending the deployment are encouraged to contact the authors. Subject to testbed-owner approval and the removal of security-sensitive information, relevant implementation guidance, code, and sanitized configuration materials may be shared to support follow-on research.

\section{Conclusion} \label{sect:s6}

This work establishes Federated Cybersecurity Testbed as a Service (FCTaaS) as a framework for coordinated cybersecurity experimentation across geographically distributed and independently managed cyber-physical system (CPS) testbeds. The framework contributes: (i) policy-aware onboarding and access control for heterogeneous testbeds; (ii) experiment lifecycle coordination across CPS assets, cloud-hosted control services, attacker environments, IDS/IPS components, and monitoring services; and (iii) a federated experimental workflow that supports the observation of attack traffic, security alerts, control communication, and operational system behavior.

FCTaaS integrates Testbed Managers, REST-based services, access-policy configurations, Kafka-based coordination, and experiment-specific VPN connectivity to support testbed discovery, authorization, initialization, monitoring, and teardown. These capabilities enable researchers to combine specialized infrastructure that would otherwise remain geographically isolated or underutilized. Such integration is particularly valuable for CPS-security studies, where meaningful experimentation often requires coordinated interaction among physical processes, industrial control interfaces, attack sources, security controls, and monitoring systems. The case studies demonstrate the practical feasibility of the proposed framework across availability- and integrity-oriented CPS-security workflows. The DoS and IDS/IPS experiments show how FCTaaS can coordinate attack generation, inline traffic inspection, packet-dropping actions, alert generation, and resource monitoring across distributed testbeds. The MITM experiment further demonstrates distributed communication interception and data injection involving a Smart Car testbed and a remote service. Collectively, these studies show that FCTaaS can support the initialization, execution, observation, and termination of heterogeneous federated CPS-security experiments under resource-intensive attack conditions.

The results also identify important limitations of federated CPS-security experimentation. Communication delay, deployment-specific network conditions, and heterogeneous endpoint resources can affect controllability, repeatability, and fidelity, particularly for timing-sensitive workflows. The current evaluation provides deployment-oriented evidence across the evaluated configurations. It does not establish performance bounds for large numbers of simultaneous users, experiments, or participating testbeds. Similarly, the reported resource measurements characterize the evaluated deployments and should not be interpreted as deployment-independent scalability guarantees. Several research directions follow from this work. First, future studies should evaluate concurrent multi-user and multi-experiment workloads, larger federations, Kafka broker behavior under higher coordination loads, VPN-management scalability, and synchronized role-aware resource profiling across heterogeneous testbeds. Second, latency-aware scheduling, adaptive coordination-plane configuration, and timing-sensitive networking mechanisms may improve support for control-critical CPS experiments. Time-Sensitive Networking represents one potential direction for reducing communication-related uncertainty in appropriate deployment environments \mbox{\cite{kopetz2022real}}. Adaptive Kafka parameter configuration may also support workload-specific tradeoffs between coordination latency and throughput \mbox{\cite{henning2021theodolite}}. Third, future extensions should strengthen federation security through improved credential lifecycle management, audit logging, endpoint attestation, fine-grained data-access controls, and systematic evaluation of threats targeting management interfaces, VPN endpoints, and participating testbeds. Broader threat coverage, including ransomware, insider threats, zero-day exploitation, and multi-stage persistent campaigns, will further expand the range of CPS-security workflows supported by FCTaaS.

Overall, FCTaaS provides a foundation for policy-aware, observable, and coordinated cybersecurity experimentation across distributed CPS infrastructure. The framework advances the use of specialized testbeds from isolated experimental assets toward composable research infrastructure for cyber-physical security evaluation.

\acknowledgement{None.}

\funding{This work was partly supported by the National Science Foundation (NSF) awards NSF-2213634, NSF-2335046, (NSF) OIA-2218046, the AGILITY project 4263090, sponsored by Korea Institute for Advancement of Technology (KIAT South Korea), and the University of Arizona’s Research, Innovation \& Impact (RII) award for the “Future Factory”.}

\authorcontributions{The authors confirm contribution to the paper as follows: conceptualization, visualization, resources, paper writing, review \& supervision---Pratik Satam; conceptualization, visualization, resources \& supervision---Salim Hariri; paper writing, paper review, experimentation \& supervision---Sicong Shao; paper writing, review \& experimentation---Ibrahim Almazyad, Josh Dean, Shalaka Satam, Qinxuan Shi, John Paul Martin Encinas, Yu-Zheng Lin, Zhanglong Yang, Tingjue Lei,  Jielun Zhang. All authors reviewed and approved the final version of the manuscript.}

\availabilityofdataandmaterials {The data that support the findings of this study are available from the Corresponding Authors, Pratik Satam and Sicong Shao, upon reasonable request.
}

\ethicsapproval{Not applicable.}

\conflictsofinterest{The authors declare no conflicts of interest.} 

\reftitle{References}



\begin{thebibliography}{10}

\bibitem{Hibler2008Large-scaleTestbed}
Hibler M, Ricci R, Stoller L, Duerig J, Guruprasad S, Stack T, et~al.
\newblock {Large-scale virtualization in the emulab network testbed}.
\newblock In: Proceedings of the USENIX Annual Technical Conference; 2008 Jun 22--27; Boston, MA, USA. p.~255--70.

\bibitem{Garcia2010}
Garcia AP, Siaterlis C, Masera M.
\newblock {Testing the fidelity of an Emulab testbed}.
\newblock In: Proceedings of the International Conference on Distributed Computing Systems; 2010 Jun 21--25; p. 307--12.

\bibitem{Berman2014GENI:Experiments}
Berman M, Chase JS, Landweber L, Nakao A, Ott M, Raychaudhuri D, et~al.
\newblock {GENI: A federated testbed for innovative network experiments}.
\newblock Comput Netw. 2014;61:5--23.

\bibitem{Goodfellow2018Merge:Ecosystems}
Goodfellow R, Thurlow L, Ravi S.
\newblock Merge: An architecture for interconnected testbed ecosystems.
\newblock arXiv:1810.08260. 2018.

\bibitem{Baldin2018TheInfrastructure}
Baldin I, Wolf T.
\newblock The future of CISE distributed research infrastructure.
\newblock ACM SIGCOMM Comput Commun Rev. 2018;48(2):46--51.

\bibitem{Friedman2019TheSystem}
Friedman T, McGeer R, Senel BC, Hemmings M, Ricart G.
\newblock {The EdgeNet system}.
\newblock In: Proceedings of the 2019 IEEE 27th International Conference on Network Protocols
  (ICNP); 2019 Oct 7--10; Chicago, IL, USA. p. 1--2.

\bibitem{Bavier2016Planetignite:Cloud}
Bavier A, McGeer R, Ricart G.
\newblock {Planetignite: A self-assembling, lightweight,
  infrastructure-as-a-service edge cloud}.
\newblock In: Proceedings of the 2016 28th International Teletraffic Congress (ITC 28); 2016 Sep 12--16; W\"urzburg, Germany. p. 130--8.

\bibitem{duplyakin2019design}
Duplyakin D, Ricci R, Maricq A, Wong G, Duerig J, Eide E, et~al.
\newblock The design and operation of CloudLab.
\newblock In: Proceedings of the 2019 USENIX Annual Technical Conference (USENIX ATC 19); 2019 Jul 10--12; Renton, WA, USA. p. 1--14.

\bibitem{White+:osdi02}
White B, Lepreau J, Stoller L, Ricci R, Guruprasad S, Newbold M, et~al.
\newblock An integrated experimental environment for distributed systems and
  networks.
\newblock In: Proceedings of the Fifth Symposium on Operating Systems Design
  and Implementation. Boston, MA, USA: {USENIX} {Association}; 2002. p. 255--70.

\bibitem{University_of_Utah_2022}
University\_of\_Utah. CloudLab hardware; 2022 [cited 2022 Feb 26].
\newblock Available from: \url{https://www.cloudlab.us/hardware.php}.

\bibitem{Fed4FIRE}
Fed4FIRE+\_Consortium. {Fed4FIRE+ Federation Framework}. [cited 2024 Feb 7].
\newblock Available from: \url{https://www.fed4fire.eu/}.

\bibitem{vandenberghe2013architecture}
Vandenberghe W, Vermeulen B, Demeester P, Willner A, Papavassiliou S, Gavras A,
  et~al.
\newblock Architecture for the heterogeneous federation of future internet
  experimentation facilities.
\newblock In: Proceedings of the 2013 Future Network \& Mobile Summit; 2013 Jul 3--5; Lisbon, Portugal. p. 1--11.

\bibitem{yala2019testbed}
Yala L, Iordache M, Bousselmi A, Imadali S.
\newblock Testbed federation for 5G experimentation: Review and guidelines.
\newblock In: Proceedings of the 2019 IEEE Conference on Standards for Communications and
  Networking (CSCN); 2019 Oct 28--30; Granada, Spain. p. 1--6.

\bibitem{gaglianese2022lightweight}
Gaglianese M, Forti S, Paganelli F, Brogi A.
\newblock Lightweight self-adaptive cloud-iot monitoring across fed4fire+
  testbeds.
\newblock In: Proceedings of the IEEE INFOCOM 2022---IEEE Conference on Computer Communications
  Workshops (INFOCOM WKSHPS); 2022 May 2--5; Virtual. p. 1--6.

\bibitem{cyber2017framework}
Group CPSPW, et~al.
\newblock Framework for cyber-physical systems: Volume 1, overview.
\newblock Gaithersburg, MD, USA: NIST Special Publication; 2017. p. 1500--201.

\bibitem{baumgartner2010virtualising}
Baumgartner T, Chatzigiannakis I, Danckwardt M, Koninis C, Kr{\"o}ller A,
  Mylonas G, et~al.
\newblock Virtualising testbeds to support large-scale reconfigurable
  experimental facilities.
\newblock In: European conference on wireless sensor networks. Berlin/Heidelberg, Germany: Springer; 2010.
  p. 210--23.

\bibitem{coulson2012flexible}
Coulson G, Porter B, Chatzigiannakis I, Koninis C, Fischer S, Pfisterer D,
  et~al.
\newblock Flexible experimentation in wireless sensor networks.
\newblock Commun ACM. 2012;55(1):82--90.

\bibitem{tunc2015claas}
Tunc C, Hariri S, Montero FDLP, Fargo F, Satam P.
\newblock CLaaS: Cybersecurity Lab as a Service--design, analysis, and
  evaluation.
\newblock In: Proceedings of the 2015 International Conference on Cloud and Autonomic Computing; 2015 Sep 21--25; Cambridge, MA, USA. p. 224--7.

\bibitem{benzel2011science}
Benzel T.
\newblock The science of cyber security experimentation: The DETER project.
\newblock In: Proceedings of the 27th Annual Computer Security Applications
  Conference; 2011 Dec 5--9; Orlando, FL, USA. p. 137--48.

\bibitem{zhou2018review}
Zhou X, Gou X, Huang T, Yang S.
\newblock Review on testing of cyber physical systems: Methods and testbeds.
\newblock IEEE Access. 2018;6:52179--94.

\bibitem{chen2014implementing}
Chen B, Butler-Purry KL, Goulart A, Kundur D.
\newblock Implementing a real-time cyber-physical system test bed in RTDS and
  OPNET.
\newblock In: Proceedings of the 2014 North American Power Symposium (NAPS); 2014 Sep 7--9; Pullman, WA, USA. p. 1--6.

\bibitem{ricci2012designing}
Ricci R, Duerig J, Stoller L, Wong G, Chikkulapelly S, Seok W.
\newblock Designing a federated testbed as a distributed system.
\newblock In: International Conference on Testbeds and Research Infrastructures. Berlin/Heidelberg, Germany: Springer; 2012. p. 321--37.

\bibitem{agarwal2016unified}
Agarwal R, Fernandez DG, Elsaleh T, Gyrard A, Lanza J, Sanchez L, et~al.
\newblock Unified IoT ontology to enable interoperability and federation of
  testbeds.
\newblock In: Proceedings of the 2016 IEEE 3rd World Forum on Internet of Things (WF-IoT); 2016 Dec 12--14; Reston, VA, USA. p. 70--5.

\bibitem{song2019ieee}
Song EY, Burns M, Pandey A, Roth T.
\newblock IEEE 1451 smart sensor digital twin federation for IoT/CPS research.
\newblock In: Proceedings of the 2019 IEEE sensors applications symposium (SAS); 2019 Mar 11--13; Sophia Antipolis, France. p.~1--6.

\bibitem{burns2018universal}
Burns M, Roth T, Griffor E, Boynton P, Sztipanovits J, Neema H.
\newblock Universal CPS environment for federation (UCEF).
\newblock In: Proceedings of the 2018 Winter Simulation Innovation Workshop; 2018 Jan 21--26; Orlando, FL, USA.

\bibitem{roth2017cyber}
Roth T, Song E, Burns M, Neema H, Emfinger W, Sztipanovits J.
\newblock Cyber-physical system development environment for energy applications.
\newblock Energy Sustain Am Soc Mech Eng. 2017;57595:V001T10A002.

\bibitem{weiss2018timing}
Weiss MA, Li-Baboud Y, Anand D, Brady KG, Boynton PA, Nguyen CT, et~al.
\newblock A calibration of timing accuracy in the NIST cyber-physical systems testbed.
\newblock In: Proceedings of the 2018 IEEE International Symposium on Precision Clock Synchronization for Measurement, Control, and Communication (ISPCS); 2018 Sep 30--Oct 5; Geneva, Switzerland.

\bibitem{Craggs2019ATestbeds}
Craggs B, Rashid A, Hankin C, Antrobus R, {\c{S}}erban O, Thapen N.
\newblock A reference architecture for IIoT and industrial control systems  testbeds.
\newblock In: Proceedings of the Living in the Internet of Things (IoT 2019); 2019 May 1--2; London, UK. p. 1--8.

\bibitem{KeaheyLessonsTestbed}
Keahey K, Anderson J, Zhen Z, Riteau P, Ruth P, Stanzione D, et~al.
\newblock Lessons learned from the chameleon testbed.
\newblock In: Proceedings of the 2020 USENIX Annual Technical Conference (USENIX ATC 20); 2020 Jul 15--17; Virtual. p.~219--33.

\bibitem{Coyne2013ABACManagement}
Coyne E, Weil TR.
\newblock {ABAC and RBAC: Scalable, flexible, and auditable access management}.
\newblock IT Professional. 2013;15(03):14--6.

\bibitem{ApacheKafka}
Kafka A. {Apache Kafka}. [cited 2022 Feb 27].
\newblock Available from: \url{https://kafka.apache.org/}.

\bibitem{Esposito2013InterconnectingService}
Esposito C, Ficco M, Palmieri F, Castiglione A.
\newblock Interconnecting federated clouds by using publish-subscribe service.
\newblock Clust Comput. 2013;16(4):887--903.

\bibitem{hussain2014threat}
Hussain S, Kamal A, Ahmad S, Rasool G, Iqbal S.
\newblock Threat modelling methodologies: A survey.
\newblock Sci Int. 2014;26(4):1607--9.

\bibitem{Pacheco2017IoTSystem}
Pacheco J, Ibarra D, Vijay A, Hariri S.
\newblock {IoT security framework for smart water system}.
\newblock In: Proceedings of the 2017 IEEE/ACS 14th International Conference on Computer Systems
  and Applications (AICCSA); 2017 Oct 30--Nov 3; Hammamet, Tunisia. p. 1285--92.

\bibitem{THEOpenplcproject.com}
OpenPLC. THE OPENPLC PROJECT. [cited 2020 Feb 2].
\newblock Available from: \url{{https://www.openplcproject.com/}}.

\bibitem{DaneelsA.1999WHATSCADA}
{Daneels A}, {Salter W}.
\newblock {What is SCADA?}
\newblock In: Proceedings of the International Conference on Accelerator and Large Experimental
  Physics Control Systems; 1999 Oct 4--8; Trieste, Italy.

\bibitem{ScadaBR}
ScadaBR. [cited 2020 Feb 27].
\newblock Available from: \url{https://sourceforge.net/projects/scadabr/}.

\bibitem{EmpoweringDocker}
Docker. {Empowering App Development for Developers|Docker}. [cited 2021 Nov 11].
\newblock Available from: \url{https://www.docker.com/}.

\bibitem{Vuletic2018REALIZATIONLINUX}
Vuleti{\'{c}} DV, Nojkovi{\'{c}} ND.
\newblock {Realization of a TCP Syn flood attack using Kali Linux}.
\newblock Mil Tech Cour. 2018;66:3.
\url{https://doi.org/10.5937/vojtehg66-16419}.

\bibitem{Huitsing2008AttackProtocols}
Huitsing P, Chandia R, Papa M, Shenoi S.
\newblock {Attack taxonomies for the Modbus protocols}.
\newblock Int J Crit Infrastruct Prot.
  2008;1:37--44.

\bibitem{Suricata}
Suricata. [cited 2020 Feb 27].
\newblock Available from: \url{https://suricata.io/}.

\bibitem{CloudAWS}
Services AW. {Cloud Services---Amazon Web Services (AWS)}. [cited 2021 Nov 11].
\newblock Available from: \url{https://aws.amazon.com/}.

\bibitem{GangliaSystem}
Ganglia. {Ganglia monitoring system}. [cited 2021 Nov 11].
\newblock Available from: \url{http://ganglia.sourceforge.net/}.

\bibitem{WenPlug-N-Pwned:IoT}
Wen H, Chen QA, Lin Z.
\newblock Plug-N-Pwned: Comprehensive vulnerability analysis of OBD-II dongles
  as a new over-the-air attack surface in automotive IoT.
\newblock In: Proceedings of the 29th USENIX Conference on Security Symposium; 2020 Aug 12--14; Virtual.

\bibitem{Sheldon2012TheNetworks}
Sheldon FT, Weber JM, Yoo SM, Pan WD.
\newblock {The Insecurity of Wireless Networks}.
\newblock IEEE Secur Priv Mag. 2012;10(4):54--61.
\newblock Available from: \url{http://ieeexplore.ieee.org/document/6193090/}.

\bibitem{Aircrack-ngAireplay-ng}
darkAudax. Aircrack-ng Getting Started. [cited 2022 Aug 17].
\newblock Available from:
  \url{https://www.aircrack-ng.org/doku.php?id=aireplay-ng}.

\bibitem{satam2020wids}
Satam P, Hariri S.
\newblock WIDS: An anomaly based intrusion detection system for Wi-Fi (IEEE
  802.11) protocol.
\newblock IEEE Trans Netw Serv Manag.
  2020;18(1):1077--91.

\bibitem{BETTERCAP}
BetterCap. [cited 2022 Feb 8].
\newblock Available from: \url{https://www.bettercap.org/}.

\bibitem{kopetz2022real}
Kopetz H, Steiner W.
\newblock Real-time communication.
\newblock In: Real-time systems: Design principles for distributed embedded
  applications. Berlin/Heidelberg, Germany: Springer; 2022. p. 177--200.

\bibitem{henning2021theodolite}
Henning S, Hasselbring W.
\newblock Theodolite: Scalability benchmarking of distributed stream processing
  engines in microservice architectures.
\newblock Big Data Res. 2021;25:100209.

\bibitem{peterson2006planetlab}
Peterson L, Muir S, Roscoe T, Klingaman A.
\newblock PlanetLab architecture: An overview.
\newblock PlanetLab Consortium; 2006.
\newblock Technical Report PDN-06-031.

\bibitem{onelab_sfa}
ONELAB.
\newblock SFA: Slice-based Federation Architecture.
\newblock 2015 [cited 2026 Aug 19].
\newblock Available from: \url{https://github.com/onelab-eu/sfa}.

\end{thebibliography}
\end{document}